\documentclass[11pt]{article}
\usepackage{axodraw}
\usepackage{epsfig}
\usepackage{amsfonts}
\usepackage{amsmath}
\usepackage{bm,bbm}
\usepackage{cite}
\usepackage{jheppub}
\allowdisplaybreaks[1]

\input pix.sty

\newcommand{\nref}[1]{(\ref{#1})}  
\newcommand{\kT}{k_\rmii{$T$}}
\newcommand{\rephase}{\phi^{+}_{(a)\I\J}}        
\newcommand{\imphase}{\phi^{-}_{(a)\I\J}}  

\newcommand{\diag}{\mathop{\mbox{diag}}}
\newcommand{\f}{f}  
\newcommand{\sH}{\rmii{$H$}}
\newcommand{\I}{\rmii{$I$}}
\newcommand{\J}{\rmii{$J$}}
\newcommand{\sL}{\rmii{$L$}}

\newcommand{\T}{\rmii{$T$}}

\newcommand{\YB}{Y_\rmii{$B$}}

\newcommand{\mW}{m_\rmii{$W$}}

\newcommand{\aL}{a^{ }_\rmii{L}}
\newcommand{\aR}{a^{ }_\rmii{R}}
\renewcommand{\eq}{eq.~}
\renewcommand{\eqs}{eqs.~}
\renewcommand{\se}{sec.~}

\renewcommand{\fig}{fig.~}
\renewcommand{\figs}{figs.~}

\newcommand{\Nc}{N_{\rm c}}

\newcommand{\bmu}{\bar\mu}

\def\lsi{\raise0.3ex\hbox{$<$\kern-0.75em\raise-1.1ex\hbox{$\sim$}}}
\def\gsi{\raise0.3ex\hbox{$>$\kern-0.75em\raise-1.1ex\hbox{$\sim$}}}
\newcommand{\lsim}{\mathop{\lsi}}
\newcommand{\gsim}{\mathop{\gsi}}

\newcommand{\nF}{n_\rmii{F}}

\newcommand{\rmii}[1]{{\mbox{\tiny\rm{#1}}}}

\newcommand{\re}{\mathop{\mbox{Re}}}
\newcommand{\im}{\mathop{\mbox{Im}}}

\newcommand{\Tint}[1]{{\hbox{$\sum$}\!\!\!\!\!\!\!\int\,}_{\!\!\!\!\raise-0.9ex\hbox{$\scriptstyle{#1}$}}}
\newcommand{\Tinti}[1]{{{\Sigma}\!\!\!\!\raise0.3ex\hbox{$\int$}_\rmii{${#1}$}}}

\newcommand{\bi}{\begin{itemize}}
\newcommand{\ei}{\end{itemize}}
\newcommand{\hide}[1]{ }

\def\TAsc(#1,#2)(#3,#4,#5)%
{\SetWidth{2.0}\CArc(#1,#2)(#3,#4,#5)\SetWidth{1.0}}
\def\Lwidth{3}

\def\TAgl(#1,#2)(#3,#4,#5){\SetWidth{2.0}\PhotonArc(#1,#2)(#3,#4,#5){\Lwidth}%
{6.283 #3 mul 360 div #4 #5 sub #4 #5 sub mul sqrt mul Tdensity mul}%
\SetWidth{1.0}}
\def\TLgl(#1,#2)(#3,#4){\SetWidth{2.0}\Photon(#1,#2)(#3,#4){\Lwidth}
{#1 #3 sub #1 #3 sub mul #2 #4 sub #2 #4 sub mul add sqrt Tdensity mul}%
\SetWidth{1.0}}

\renewcommand{\picc}[1]{\;\parbox[c]{60pt}{\begin{picture}(60,30)(0,0)
\SetWidth{1.0}\SetScale{1.3} #1 \end{picture}}\; }
\def\Lwidth{1.3}

\renewcommand{\pic}[1]{\;\parbox[c]{30pt}{\begin{picture}(30,30)(0,-3)
\SetWidth{1.0}\SetScale{0.8} #1 \end{picture}}\;}
\renewcommand{\picb}[1]{\;\parbox[c]{45pt}{\begin{picture}(45,30)(0,-3)
\SetWidth{1.0}\SetScale{0.8} #1 \end{picture}}\;}

\def\filledsquare{\parbox[c]{5pt}{%
 \begin{picture}(5,5)(0,0)
 \SetWidth{1.0}\SetScale{1.0}
 \GBox(0,0)(5,5){0}
 \end{picture}}\,
}

\makeatletter \@addtoreset{equation}{section} \makeatother
\renewcommand{\theequation}{\arabic{section}.\arabic{equation}}
\makeatletter
\renewcommand\section{\@startsection {section}{1}{\z@}%
                                   {-5.5ex \@plus -1ex \@minus -.2ex}
                                   {2.3ex \@plus.2ex}%
                                   {\normalfont\large\bfseries}}
\renewcommand\subsection{\@startsection{subsection}{2}{\z@}%
                                     {-3.25ex\@plus -1ex \@minus -.2ex}%
                                     {1.5ex \@plus .2ex}%
                                     {\normalfont\normalsize\bfseries}}
\renewcommand\thesection {\@arabic\c@section}
\renewcommand\thesubsection   {\thesection.\@arabic\c@subsection}
\renewcommand{\@seccntformat}[1]{%
\csname the#1\endcsname.\hspace{1.0em}}
\makeatother

\begin{document}

\flushbottom


\begin{flushright}
CERN-TH-2019-062\\ 
July 2019
\end{flushright}

\vspace*{0.5cm}

\title{Sterile neutrino dark matter via GeV-scale leptogenesis? }

\author[a]{J.~Ghiglieri,}
\author[b]{M.~Laine}

\affiliation[a]{%
Theoretical Physics Department, CERN, \\ 
CH-1211 Geneva 23, Switzerland}
\affiliation[b]{%
AEC, 
Institute for Theoretical Physics, 
University of Bern, \\ 
Sidlerstrasse 5, CH-3012 Bern, Switzerland}

\emailAdd{jacopo.ghiglieri@cern.ch}
\emailAdd{laine@itp.unibe.ch}

\abstract{%
It has been proposed that in a part of the parameter space 
of the Standard Model completed by three generations of keV...GeV  
right-handed neutrinos, neutrino masses, dark matter, and baryon 
asymmetry can be accounted for simultaneously. Here we numerically solve 
the evolution equations describing the cosmology of this scenario in 
a 1+2 flavour situation at temperatures $T \le 5$~GeV, 
taking as initial conditions maximal lepton asymmetries produced dynamically 
at higher temperatures, and accounting for late entropy and lepton asymmetry
production as the heavy flavours fall out of equilibrium and decay.
For 7~keV dark matter mass and other parameters 
tuned favourably, $\sim 10$\% of the observed abundance can be  
generated. Possibilities for increasing the abundance are enumerated.
}
%


 
 
\maketitle

%
\section{Introduction}
\la{se:intro} 

The idea of accounting for dark matter through 
keV-scale sterile neutrinos~\cite{dw,sf} 
is strongly constrained by now
(for a review see, e.g., ref.~\cite{review}).
The non-observation of $\gamma$-rays from 
putative sterile neutrino decays restricts 
their Yukawa couplings to be very small, $|h^{ }_{\I a}| <  10^{-12}$.
With such small couplings 
a sufficient number of sterile neutrinos can be produced in 
the Early Universe only if the production 
is enhanced through a resonant mechanism~\cite{sf}, 
requiring the presence of large lepton asymmetries. 
Some time ago, it was pointed out~\cite{singlet} 
that this scenario could be embedded
in a framework in which two generations of GeV-scale right-handed 
neutrinos first generate a baryon asymmetry~\cite{ars,as}, and then 
continue to generate lepton asymmetries, which subsequently
boost dark matter production~\cite{shifuller,aba,dmpheno}. 

In the most detailed dark matter
computation carried out so far~\cite{dmpheno}, it was assumed that 
lepton asymmetries are produced first, at $T \gsim 5$~GeV, 
whereas dark matter production is only active at $T \lsim 5$~GeV. 
However, if the mass scale of the heavier sterile neutrinos
is $M \; \gsim 2$~GeV, they decay at $T \ll M/\pi \; \lsim $~GeV, 
and these decays may
produce further lepton asymmetries~\cite{singlet,late}.  
Then lepton asymmetry generation and dark matter production may proceed
simultaneously, and need to be accounted for within a unified framework. 

The purpose of the present paper is to assume that the initial 
lepton asymmetries have 
been dynamically produced by two generations of 
GeV-scale right-handed neutrinos. In a recent work~\cite{degenerate}, 
we showed 
that in this case lepton asymmetries $\gsim 10^3$ times larger than
the baryon asymmetry can arise. Furthermore, the lepton
asymmetries have an intriguing structure, being evenly distributed
amongst all flavours and settling into a stationary state
(see also ref.~\cite{eijima}). 
We now follow that state down to lower temperatures, 
at which the GeV-scale right-handed
neutrinos freeze out and decay. This non-equilibrium dynamics
modifies the expansion rate of the universe and may also produce
new lepton asymmetries. 
The question is whether this
could help to boost the asymmetries that, according to 
ref.~\cite{degenerate}, were too small to have a substantial
effect in the dark matter context. For the dark matter
sector itself we fix the mass to the prototypical 7 keV scenario, 
with the corresponding Yukawa couplings pushed to the maximal allowed 
range as suggested by supposed 
observations~\cite{observe1,observe2}.\footnote{%
 At the time of writing 
 these observations continue to be controversially discussed.
 }

The presentation is organized as follows.
The rate equations applying to the 1+2 sterile neutrino
system are summarized in \se\ref{se:review}. In \se\ref{se:expansion}
we explain how the falling out of equilibrium 
of the ``heavy'' GeV-scale flavours 
modifies the expansion of the universe, and transcribe the 
rate equations to this situation. Subsequently, 
the heavy part of the rate equations 
can be simplified as explained in \se\ref{se:heavy},
whereas the ``light'' keV-scale part may experience
resonant enhancement, cf.\ \se\ref{se:light}, which prohibits 
any substantial simplification. Parameter choices are 
justified in \se\ref{se:params}, and numerical results are 
presented in \se\ref{se:abundance}. 
A brief summary and outlook conclude this investigation
in \se\ref{se:summary}.

%
\section{Review of rate equations for the 1+2 flavour situation}
\la{se:review}

The theory we work with is described by the Lagrangian
\be
 \mathcal{L} \; = \; 
 \mathcal{L}^{ }_\rmii{SM} 
 + 
  \fr12 \bar{N}^{ }_{\!\I} 
 \bigl( i \gamma^\mu\partial_\mu - M^{ }_{\!\I} \bigr) {N}^{ }_{\!\I}
 - 
 \Bigl(
  \bar{\ell}^{ }_{a}  \aR \tilde{\phi}\, h^{*}_{\I a}\, N^{ }_{\!\I}
  +
  \bar{N}^{ }_{\!\I} \, h^{ }_{\I a}\, \tilde{\phi}^\dagger  \aL \ell^{ }_a
 \Bigr)
 \;, \la{L}
\ee
where $M^{ }_\I \ge 0$ are Majorana masses; 
$\tilde{\phi} = i \sigma_2 \phi^*$ is a Higgs doublet; 
$\aL, \aR$ are chiral projectors;  $\ell^{ }_a = (\nu\, e)^T_a$ 
is a left-handed lepton doublet of generation $a$; 
$h^{ }_{\I a}$ are the components of the neutrino Yukawa matrix; 
and summations over indices are left implicit.

We consider the situation 
$
 M^{ }_1 \sim \mbox{keV} \ll M^{ }_{2,3} \sim \mbox{GeV}
$, 
and assume the 2nd and 3rd generations to be almost degenerate 
in mass. The average mass is denoted by
$M^{ }_\sH \equiv (M^{ }_2 + M^{ }_3)/2$. 
The heavy flavours $I = 2,3$ and the 
associated Yukawa couplings $|h^{ }_{\I a}| \lsim 10^{-7}$ are chosen to 
reproduce the active neutrino mass differences and mixing angles, 
whereas the first generation has much smaller 
Yukawa couplings, $|h^{ }_{1a}| <  10^{-12}$,  
as is suitable for playing a role in dark matter physics. 

The density matrix of the hierarchical 1+2 flavour system is expressed as
\be
 \rho^{\pm}_{ }
 \; \equiv \; 
 \left( 
  \begin{array}{cc}
    \f^{\pm}_{ } & 0 \\   
     0  & \rho^{\pm}_\sH 
  \end{array}
 \right)
 \;, \quad
 \rho^{\pm}_{\I\J} 
 \; \equiv \; 
 \bigl\{ \rho^{\pm}_{\sH} \bigr\}^{ }_{\I\J}
 \;, \quad
 I,J \in \{2,3\}
 \;, \la{def_rho_block}
\ee
and similarly for other objects. Here
\be
 f^{\pm}_{ }
 \; \equiv \; 
 \frac{ f^{ }_\rmii{$(+)$} \pm f^{ }_\rmii{$(-)$} }{2}
 \;, \quad
 \rho^{\pm}_{ } 
 \; \equiv \; 
 \frac{\rho^{ }_\rmii{$(+)$} \pm \rho^{ }_\rmii{$(-)$}}{2}
 \la{symmetrization} 
\ee
denotes a symmetrization/antisymmetrization
with respect to helicity $(\pm)$, and off-diagonal heavy-light components 
of $\rho^{\pm}_{ }$ have averaged out up to effects 
suppressed by $1/M^{ }_{\sH}$. 
The lepton asymmetry in flavour $a$ 
is denoted by $n^{ }_a$,
whereas $n^{ }_\rmii{$B$}$ is the baryon asymmetry.
 
The evolution equation for lepton 
asymmetries can be split into the contributions
of the light and heavy flavours,\footnote{%
 In order to derive the evolution equations
 \nref{d_na_1p2}--\nref{d_rhoH_1p2}, we have generalized
 the considerations in refs.~\cite{sr,dmpheno,selfE,cptheory}
 to apply to three flavours of sterile neutrinos possessing 
 an arbitrary mass spectrum, and at the end simplified the setup
 by specializing to a hierarchical 1+2-flavour system.  
 } 
\ba
 \dot{n}^{ }_a - \frac{\dot{n}^{ }_\rmii{$B$}}{3}
 & = & 
 4 \int_{\vec{k}} \Bigl\{ 
   \bigl[f^{+}_{ } - \nF^{ }(\omega^{ }_1) \bigr]
   \, {B}^{+}_{(a)11} 
 + f^{-}_{ } {B}^{-}_{(a)11}
 - \nF^{ }(\omega^{ }_1) \bigl[1 - \nF^{ }(\omega^{ }_1)\bigr]\, 
  {A}^+_{(a)11}
 \Bigl\} 
 \la{d_na_1p2} \\ 
 & + &   
 4 \int_{\vec{k}} \tr \Bigl\{ 
   \bigl[\rho^{+}_{\sH} - \nF^{ }(\omega^{ }_\sH) \bigr]
    {B}^{+}_{(a)\sH} 
 + \rho^{-}_{\sH}\, {B}^{-}_{(a)\sH}
 - \nF^{ }(\omega^{ }_\sH) \bigl[1 - \nF^{ }(\omega^{ }_\sH)\bigr]\,
  {A}^+_{(a)\sH}
 \Bigr\} \hspace*{4mm}
 \;, \nonumber 
\ea
where $\nF^{ }$ denotes the Fermi distribution, 
$
 \int_{\vec{k}} \equiv \int 
 \! \frac{{\rm d}^3\vec{k}}{(2\pi)^3}
$, 
and $\omega^{ }_{\I} \equiv \sqrt{k^2 + M_\I^2}$.
The light and heavy components of the density matrix evolve as 
\ba
 \dot{\f}^{\pm} 
 & = & 
 2 {D}^{\pm}_{11}\, \bigl[ \nF^{ }(\omega^{ }_1) - \f^{+}_{ } \bigr]
 - 
 2 {D}^{\mp}_{11}\, \f^{-}_{ }
 + 
 2 {C}^{\pm}_{11}\,
 \nF^{ }(\omega^{ }_1) \bigl[ 1 - \nF^{ }(\omega^{ }_1) \bigr] 
 \;, \la{d_rhoL_1p2}
 \\[2mm] 
 \dot{\rho}^{\pm}_\sH & = & 
   i \bigl[
   \diag({\omega}^{ }_2,{\omega}^{ }_3)
    - {H}^{+}_{\!\sH}, \rho^{\pm}_{\sH}
   \bigr]
 \; - \;  
   i \bigl[
     {H}^{-}_{\!\sH}, \rho^{\mp}_{\sH}
   \bigr]
 \nn[2mm] 
 & + & 
 \bigl\{  
   {D}^{\pm}_{\sH} \,,\,
   \nF^{ }(\omega^{ }_{\sH}) - \rho^+_{\sH}
 \bigr\}
 \; - \;  
 \bigl\{  
   {D}^{\mp}_{\sH} \,,\,
   \rho^-_{\sH}
 \bigr\}
 \; + \; 
 2 {C}^{\pm}_{\sH}\, \nF^{ }(\omega^{ }_{\sH})
 \bigl[ 1 - \nF^{ }(\omega^{ }_{\sH}) \bigr]
 \;. \la{d_rhoH_1p2}  
\ea
The coefficients associated with the light flavours read
\ba
 && \hspace*{-1.3cm}
 {A}^{+}_{(a)11} 
 \; = \; 
 \bmu^{ }_a \phi^{+}_{(a)11} \, {Q}^{+}_{(a)\sL}
 \;, \quad 
 {B}^{+}_{(a)11} 
 \; = \; 
 \phi^{+}_{(a)11} \,{\!\bar{Q}}^{+}_{(a)\sL}
 \;, \quad
 {B}^{-}_{(a)11} 
 \; = \; 
 \phi^{+}_{(a)11} \, {Q}^{-}_{(a)\sL}
 \;, \la{B_11} \\
 && \hspace*{-1.3cm}
 {C}^{+}_{11} 
 \; = \; 
 {\textstyle\sum_a}\, \bmu^{ }_a\, \phi^{+}_{(a)11} \,
 {\!\bar{Q}}^{+}_{(a)\sL}
 \;, \quad
 {C}^{-}_{11} 
 \; = \; 
 {\textstyle\sum_a}\, \bmu^{ }_a\, \phi^{+}_{(a)11} \, 
 {Q}^{-}_{(a)\sL}
 \;, \la{C_11} \\
 && \hspace*{-1.3cm}
 {D}^{+}_{11} 
 \; = \; 
 {\textstyle\sum_a}\, \phi^{+}_{(a)11} \,
 {Q}^{+}_{(a)\sL}
 \;, \quad
 {D}^{-}_{11} 
 \; = \; 
 {\textstyle\sum_a}\, \phi^{+}_{(a)11} \,
 {\!\bar{Q}}^{-}_{(a)\sL}
 \;, \la{D_11} 
\ea
where $(...)^{ }_{\sL}$ indicates the use 
of a ``light'' mass $M^{ }_1$, whereas
the heavy coefficients read
\ba
 {A}^{+}_{(a)\I\I} & = &
   \bmu^{ }_a \phi^{+}_{(a)\I\I}
      {Q}^{+}_{(a)\sH}
 \;, \quad  
 {B}^{\pm}_{(a)\I\J} \; = \; 
     \phi^{\mp}_{(a)\I\J} {Q}^{\pm}_{(a)\sH}
   + \phi^{\pm}_{(a)\I\J} \,{\!\bar{Q}}^{\pm}_{(a)\sH}
 \;, \la{matB} \\
 {C}^{\pm}_{\I\J} & = & 
      {\textstyle\sum_a}\, \bmu^{ }_a\, 
      \bigl[\,
      \phi^{\mp}_{(a)\I\J}
      {Q}^{\pm}_{(a)\sH} 
    + \phi^{\pm}_{(a)\I\J}
     \,{\!\bar{Q}}^{\pm}_{(a)\sH} 
     \,\bigr]
 \;, \la{matC} \\ 
 {D}^{\pm}_{\I\J} & = & 
      {\textstyle\sum_a} 
    \bigl[\,
    \phi^{\pm}_{(a)\I\J} 
    {Q}^{\pm}_{(a)\sH}
  + \phi^{\mp}_{(a)\I\J}\,
   {\!\bar{Q}}^{\pm}_{(a)\sH} 
    \,\bigr]
 \;, \la{matD} \\
 {H}^{\pm}_{\I\J} & = & 
     {\textstyle\sum_a} 
   \bigl[\, 
   \phi^{\pm}_{(a)\I\J} 
   {U}^{\pm}_{(a)\sH}
 + \phi^{\mp}_{(a)\I\J} 
   \,{\!\bar{U}}^{\pm}_{(a)\sH} 
   \, \bigr]
 \;, \la{matH1}  
\ea
where $(...)^{ }_{\sH}$ stands for a ``heavy'' mass $M^{ }_{\sH}$. 
We have denoted leptonic chemical 
potentials by $\bmu^{ }_a \equiv \mu^{ }_a/T$, 
and expressed the dependence on neutrino Yukawa couplings through 
\be
 \rephase \; \equiv \; \re ( h^{ }_{\I a}h^*_{\J a} )
 \;, \quad
 \imphase \; \equiv \; - i \im ( h^{ }_{\I a}h^*_{\J a} )
 \;, \la{yukawas}
\ee 
whereas
$
 {Q}^{\pm}_\rmii{$(a)$} = [{Q}^{ }_\rmii{$(a+)$}\pm
 {Q}^{ }_\rmii{$(a-)$}]/2
$ denote symmetrization and antisymmetrization with respect to helicity. 
The coefficients ${Q}$ and $\,{\!\bar{Q}}$ 
parametrize the C-even and C-odd parts, respectively, 
of ``absorptive'' reactions (i.e.\ real processes), 
whereas ${U}$ and $\,{\!\bar{U}}$ 
parametrize ``dispersive'' corrections. Specifically, 
\be
 \frac{
   \bar{u}^{ }_{\vec{k}\tau\I}
       \im \Pi^\rmii{R}_{a} (\mathcal{K}^{ }_\I)\,
     u^{ }_{\vec{k}\tau\I} 
   }
 {\omega^{ }_\I}
 \; \equiv \;  Q^{ }_{(a\tau)\I} 
 + \bar{Q}^{ }_{(a\tau)\I} 
 \;, \quad
 \frac{
   \bar{u}^{ }_{\vec{k}\tau\I}
       \re \Pi^\rmii{R}_{a} (\mathcal{K}^{ }_\I)\,
     u^{ }_{\vec{k}\tau\I} 
   }
 {\omega^{ }_\I}
 \; \equiv \;  U^{ }_{(a\tau)\I} 
 + \bar{U}^{ }_{(a\tau)\I} 
 \;, \la{QbarQ_UbarU}
\ee
where $ \Pi^\rmii{R}_a $ is a retarded correlator associated
with the operator 
$
  j^{ }_a = 
 \tilde{\phi}^\dagger \aL \ell^{ }_a
$
to which the sterile neutrinos couple;
$\tau = \pm$ denotes helicity; 
$I = L,H$ refers to the flavour; and $Q$ and $\bar{Q}$ can be extracted
by symmetrizing and antisymmetrizing in chemical potentials, respectively.  

For a practical determination of $Q,\bar{Q},U,\bar{U}$, 
we have generalized
the computations of refs.~\cite{numsm,broken,degenerate} 
to arbitrary kinematics 
(i.e.\ not only the ultrarelativistic regime $\pi T \gg M$ but
also $\pi T \sim M$ or $\pi T \ll M$), restricting however 
still to the approximation 
$M \ll \mW^{ }$ in the treatment of $2\leftrightarrow 2$
scatterings below the electroweak crossover.  

In order to close the system, the chemical potentials appearing 
in \eqs\nref{B_11}--\nref{QbarQ_UbarU} need to be re-expressed in terms
of the number densities appearing on the left-hand side of 
\eq\nref{d_na_1p2}. This requires the determination of 
``susceptibilities''. We follow the approach in appendix~A
of ref.~\cite{degenerate}, simplifying
the formulae by restricting to $T^2 \ll v^2$,
where $v\sim 246$~GeV, but 
adding charged lepton and light quark masses 
according to ref.~\cite{dmpheno}. Hadronic 
contributions are smoothly switched off at low $T$ by a replacement 
$\Nc^{ }\to N^{ }_{\rm c,eff}$, as proposed in ref.~\cite{numsm}. 

%
\section{Non-equilibrium expansion}
\la{se:expansion}

The GeV-scale flavours that are responsible 
for leptogenesis at $T \sim 130$~GeV, freeze out and subsequently decay 
when $\pi T \ll M^{ }_{\sH}$. These non-equilibrium decays 
release entropy~\cite{st}, an effect which has been argued to be substantial  
for $M^{ }_{\sH} \simeq 1...10$~GeV~\cite{entropy}, 
and which therefore needs to be included in 
dark matter and baryogenesis computations.
(When $\pi T \gg M^{ }_{\sH}$, the GeV-scale flavours already 
have a small effect on the energy and entropy densities, however
this is on the percent level and thus insignificant on our resolution.)

Denoting by $a(t)$ the cosmological scale factor and by 
$
 m^{ }_\rmi{Pl} = 1.22091 \times 10^{19}
$~GeV
the Planck mass, and assuming a flat universe, Friedmann equations
can be expressed as 
\ba
 \frac{\dot{a}}{a} & = & \sqrt{\frac{8\pi}{3}} 
                         \frac{\sqrt{e}}{m^{ }_\rmi{Pl}}
 \; \equiv \; H 
 \;, \la{einstein1} \\ 
 {\rm d}(e\, a^3) & = & - p\, {\rm d}(a^3)  
 \;, \la{einstein2}
\ea
where $e$ is the energy density, $p$ is the pressure, and
$H$ is the Hubble rate. We write
\be
 e = e^{ }_\T + e^{ }_{\sH}
 \;, \quad
 p = p^{ }_\T + p^{ }_{\sH}
 \;, \la{splitup}
\ee
where $e^{ }_\T$ and $p^{ }_\T$ are the Standard Model
energy density and pressure at a temperature~$T$, whereas
$e^{ }_{\sH}$ and $p^{ }_{\sH}$ represent the contribution of the heavy
right-handed neutrinos. 
If we denote by 
\be
 k^{ }_t \; \equiv \; k\, \frac{a(t^{ }_0)}{a(t)}
 \la{kt}
\ee
a co-moving momentum mode and by 
$
 \int^{ }_{k_t} \equiv \int\!\frac{{\rm d}^3\vec{k}^{ }_t}{(2\pi)^3}
$
the corresponding phase space integral,  
the energy density and pressure carried by the heavy flavours
can be expressed as
\be
 e^{ }_{\sH}  =  \sum_{\I} \int^{ }_{k_t}  
 2 \rho^+_{\I\I}(t,k^{ }_t) \, \omega^{ }_{\I}
 \;, \quad
 p^{ }_{\sH} =  \sum_{\I} \int^{ }_{k_t}  
 2 \rho^+_{\I\I}(t,k^{ }_t) \, \frac{k_t^2}{3\omega^{ }_{\I}}
 \;,  \quad
 \omega^{ }_{\I} \equiv \sqrt{M^2_{\I} + k_t^2}
 \;. \la{pH}
\ee 

We now insert \eq\nref{splitup} into \eq\nref{einstein2},  
and move the thermal terms to the left-hand side. Making use 
of ${\rm d} e^{ }_\T = T {\rm d} s^{ }_\T$ and 
$e^{ }_\T + p^{ }_\T = T s^{ }_\T$, 
where  $ s^{ }_\T $ is the Standard Model entropy density, 
we find
\be
 T \partial^{ }_t (s^{ }_\T \, a^3)
 =  
 - \partial^{ }_t (e^{ }_{\sH} a^3)
 - p^{ }_{\sH} \, \partial^{ }_t (a^3)
 \;, \quad
 \partial^{ }_t \; \equiv \; 
 \frac{{\rm d}}{{\rm d}t}
 \;. \la{entropy_law}
\ee
In order to proceed, it is helpful
to express the phase-space integrals 
in \eq\nref{pH}
in terms of the time-independent
variable $k$ (cf.\ \eq\nref{kt}), 
because then $\rho^+_{\I\I}$ appears in a form for
which a time-evolution equation is available.
For a combination appearing in \eq\nref{entropy_law}
this implies  
\be
 (e^{ }_{\sH} a^3)(t)
 \; = \; 
 a^3(t^{ }_0) \sum_{\I} \int^{ }_{k}
 2 \rho^+_{\I\I} \biggl( t,k\,\frac{a(t^{ }_0)}{a(t)} \biggr) 
 \, 
 \sqrt{M_{\I}^2 + k^2 \, \frac{a^2(t^{ }_0)}{a^2(t)}}
 \;. 
\ee
A derivative with respect to $t$ now operates on two terms, 
$ \rho^+_{\I\I} $ as well as the last piece, 
\be
 \partial^{ }_t
 \sqrt{M_{\I}^2 + k^2 \, \frac{a^2(t^{ }_0)}{a^2(t)}}
 = - \frac{k_t^2 H}{\omega^{ }_{\I}}
 \;. \la{d_wH}
\ee
Once $p^{ }_{\sH}$ from \eq\nref{pH} is inserted, 
the contribution from \eq\nref{d_wH}
cancels against the contribution
from $p^{ }_{\sH}$ in \eq\nref{entropy_law}. In total, then,  
\be
 T \partial^{ }_t (s^{ }_\T \, a^3) 
 = 
 - 
 a^3(t^{ }_0) \sum_{\I} \int^{ }_{k}
 2 \partial^{ }_t\,
 \rho^+_{\I\I} \biggl( t,k\,\frac{a(t^{ }_0)}{a(t)} \biggr)
 \, 
 \sqrt{M_{\I}^2 + k^2 \, \frac{a^2(t^{ }_0)}{a^2(t)}}
 \;. \la{dsT}
\ee

At this point we make use of the equation of motion of $\rho^+_{\I\I}$.
It follows from \eq\nref{d_rhoH_1p2} that, to a good approximation, 
\be
 \partial^{ }_t\, 
 \rho^+_{\I\I} \biggl( t,k\,\frac{a(t^{ }_0)}{a(t)} \biggr)
 = 
 \Gamma^{ }_{\I\I} \, 
 \bigl[\, \nF^{ }(\omega^{ }_{\I}) - \rho^+_{\I\I}(t,k^{ }_t)\, \bigr]
 \;, \quad 
 \Gamma^{ }_{\I\I}
 \; \equiv \; 
 2\, {\textstyle\sum_a} \phi^+_{(a)\I\I}\, Q^+_{(a)\sH}
 \;.  \la{rhop_eom}
\ee
Inserting \eq\nref{rhop_eom} into \eq\nref{dsT}
and going subsequently back to co-moving momenta
as integration variables, we get 
\be
 T \partial^{ }_t (s^{ }_\T a^3)
 = 
 a^3(t) \sum_{\I} \int^{ }_{k_t} 2 \omega^{ }_{\I}\, \Gamma^{ }_{\I\I}
 \bigl[\, \rho^+_{\I\I}(t,k^{ }_t) - \nF^{ }(\omega^{ }_{\I}) \,\bigr]
 \;. \la{sT_eom}
\ee
We see that entropy is generated only if $ \rho^+_{\I\I} $ falls out
of equilibrium. 

Let us simplify the setup by making use of so-called momentum averaging. 
Even though not associated
with any formally small expansion parameter, 
this turns out to represent a reasonable approximation
in many cases~\cite{degenerate,kinetic,cpnumerics}.
We integrate \eq\nref{rhop_eom} over $k$ and then change variables
into $k^{ }_t$, which leads to 
\be
 \partial^{ }_t \Bigl[ 
 a^3(t) \int^{ }_{k_t} \rho^+_{\I\I}(t,k^{ }_t)
 \Bigr]
 = 
 a^3(t) \int^{ }_{k_t} 
 \Gamma^{ }_{\I\I} \, 
 \bigl[\, \nF^{ }(\omega^{ }_{\I}) - \rho^+_{\I\I}(t,k^{ }_t)\, \bigr]
 \;. \la{rhop_ave}
\ee
Now introduce the ansatz
\be
 \rho^+_{\I\I}(t,k^{ }_t) 
 \;\simeq\; 
 \nF^{ }(\omega^{ }_{\I}) 
 \, 
 \frac{Y^+_{\I\I}(t)}{Y^+_\rmi{eq}(t)}
 \;, \quad
 Y^{+}_\rmi{eq} \; \equiv \; 
 \frac{ \int^{ }_{k_t} \nF^{ }(\omega^{ }_{\I}) }
      { s^{ }_\T }
 \;, 
\ee
where $Y^+_{\I\I}$ is a yield parameter, and denote  
\be
 \bigl\langle ... \bigr\rangle^{ }_1 \; \equiv \; 
 \frac{ \int^{ }_{k_t} (...) \nF^{ }(\omega^{ }_{\I}) }
      { \int^{ }_{k_t} \nF^{ }(\omega^{ }_{\I}) }
 \;. \la{momav_1}
\ee
Then \eqs\nref{sT_eom} and \nref{rhop_ave} become
\ba
 T \partial^{ }_t (s^{ }_\T a^3)
 & = & 
 s^{ }_\T a^3
  \, 
 \sum_{\I} 
 2 \bigl\langle \omega^{ }_{\I}\, \Gamma^{ }_{\I\I} 
 \bigr\rangle^{ }_{1}
  \, 
 \bigl(Y^+_{\I\I} - Y^+_\rmi{eq}\bigr)
 \;, \la{eom_1} \\ 
 \partial^{ }_t \bigl( Y^+_{\I\I} \, s^{ }_\T a^3 \bigr) 
 & = & 
 s^{ }_\T a^3 
  \, 
 \bigl\langle \Gamma^{ }_{\I\I} 
 \bigr\rangle^{ }_{1}
  \, 
 \bigl( Y^+_\rmi{eq} - Y^+_{\I\I} \bigr)
 \;. \la{eom_2}
\ea

As a final step, the evolutions of 
$Y^+_{\I\I}$ and 
$s^{ }_\T a^3$ can be decoupled from each other, by inserting
\eq\nref{eom_1} into \nref{eom_2}. Moreover, 
introducing 
\be
 x \; \equiv \; \ln\biggl( \frac{T^{ }_\rmi{max}}{T} \biggr)
 \;, 
 \quad
 \mathcal{J} \; \equiv \; \frac{{\rm d}x}{{\rm d}t}
 \; = \; - \frac{\dot{T}}{T}
 \;, 
\ee 
we can rewrite 
\be
 \frac{ \partial^{ }_t (s^{ }_\T a^3) }{s^{ }_\T a^3} = 
 \frac{\dot{T} s_\T'}{s^{ }_\T}  
 + \frac{3\dot{a}}{a} 
 = 
 - \frac{\mathcal{J}}{c_s^2} + 3 H
 \;, \la{eom_3}
\ee
where $c_s^2$ is the speed of sound squared. From \eqs\nref{eom_1} 
and \nref{eom_3} the Jacobian $\mathcal{J}$ can be solved for, 
\be
 \mathcal{J} = 
 c_s^2 \, \biggl\{ 
 3 H - \sum_{\I} \frac{
 2 \langle \omega^{ }_{\I}\, \Gamma^{ }_{\I\I} \rangle^{ }_1
 \, ( Y^+_{\I\I} - Y^+_\rmii{eq} ) }{T}
 \biggr\} 
 \;, \quad
 H = \sqrt{\frac{8\pi}{3}} 
 \frac{\sqrt{ e^{ }_\T 
     + s^{ }_\T \sum_{\I}
   2 \langle \omega^{ }_{\I} \rangle^{ }_1 Y^+_{\I\I} }}
 {m^{ }_\rmi{Pl}}
 \;. \la{J}
\ee
Then the basic equations become
\ba
 \partial^{ }_x Y^+_{\I\I} & = & 
 - \frac{
  \bigl\langle \Gamma^{ }_{\I\I} 
  \bigr\rangle^{ }_{1} }{\mathcal{J}} \, 
  \bigl( Y^+_{\I\I} - Y^+_\rmi{eq} \bigr)
 - 
 \partial^{ }_x \ln(s^{ }_\T a^3) \,
  Y^+_{\I\I} 
 \;, \la{xeq_1} \\ 
 \partial^{ }_x \ln(s^{ }_\T a^3)
 & = & 
 \frac{1}{\mathcal{J}} \, 
 \biggl\{
  \sum_{\J} \frac{
    2 \langle \omega^{ }_{\J} \Gamma^{ }_{\!\J\J} \rangle^{ }_1
    \, ( Y^+_{\J\J} - Y^+_\rmii{eq} ) }{T}
 \biggr\}
 \;. \la{xeq_2}
\ea

\begin{figure}[t]

\hspace*{-0.1cm}
\centerline{%
 \epsfysize=7.5cm\epsfbox{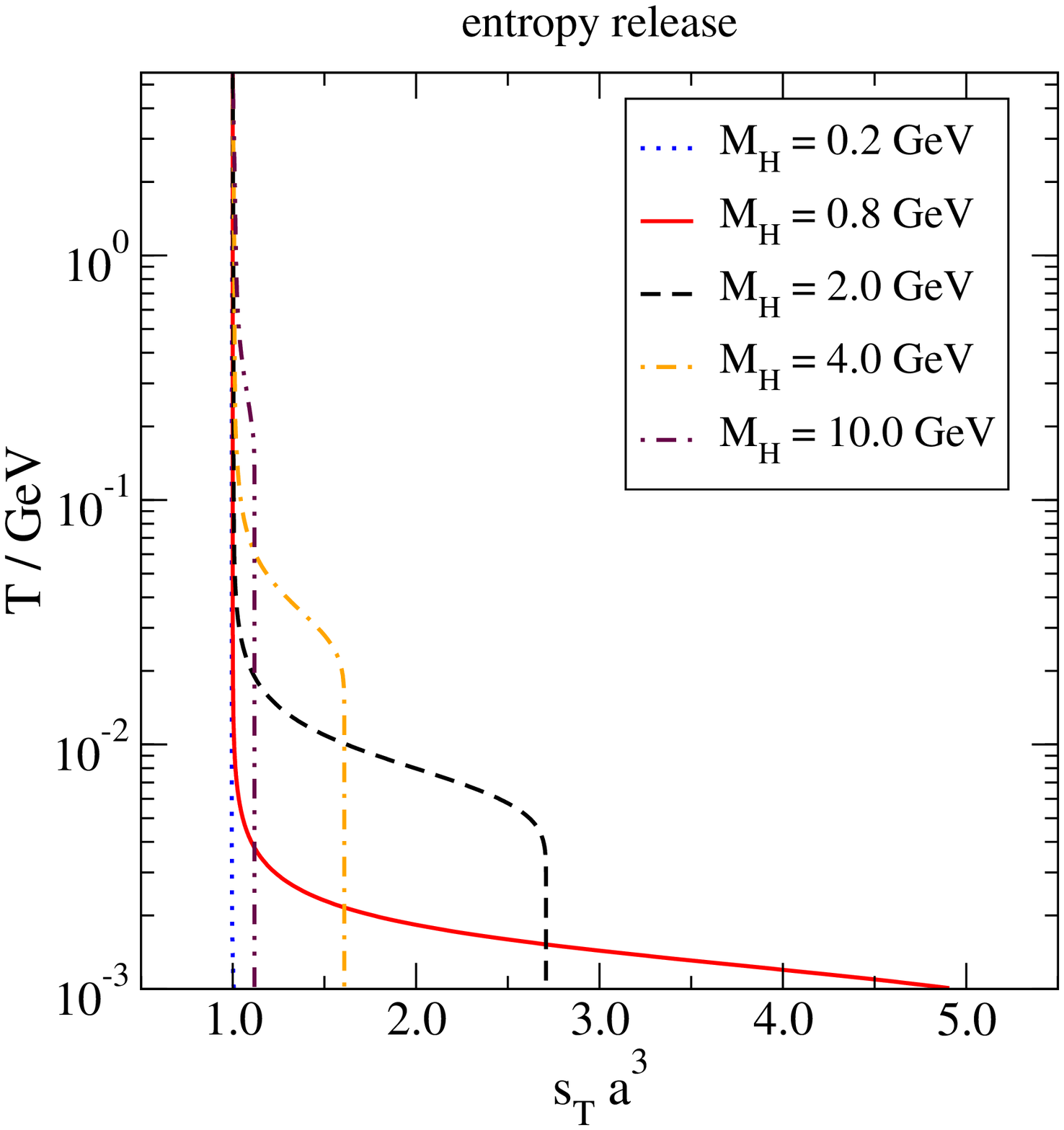}%
 \hspace{0.5cm}%
 \epsfysize=7.5cm\epsfbox{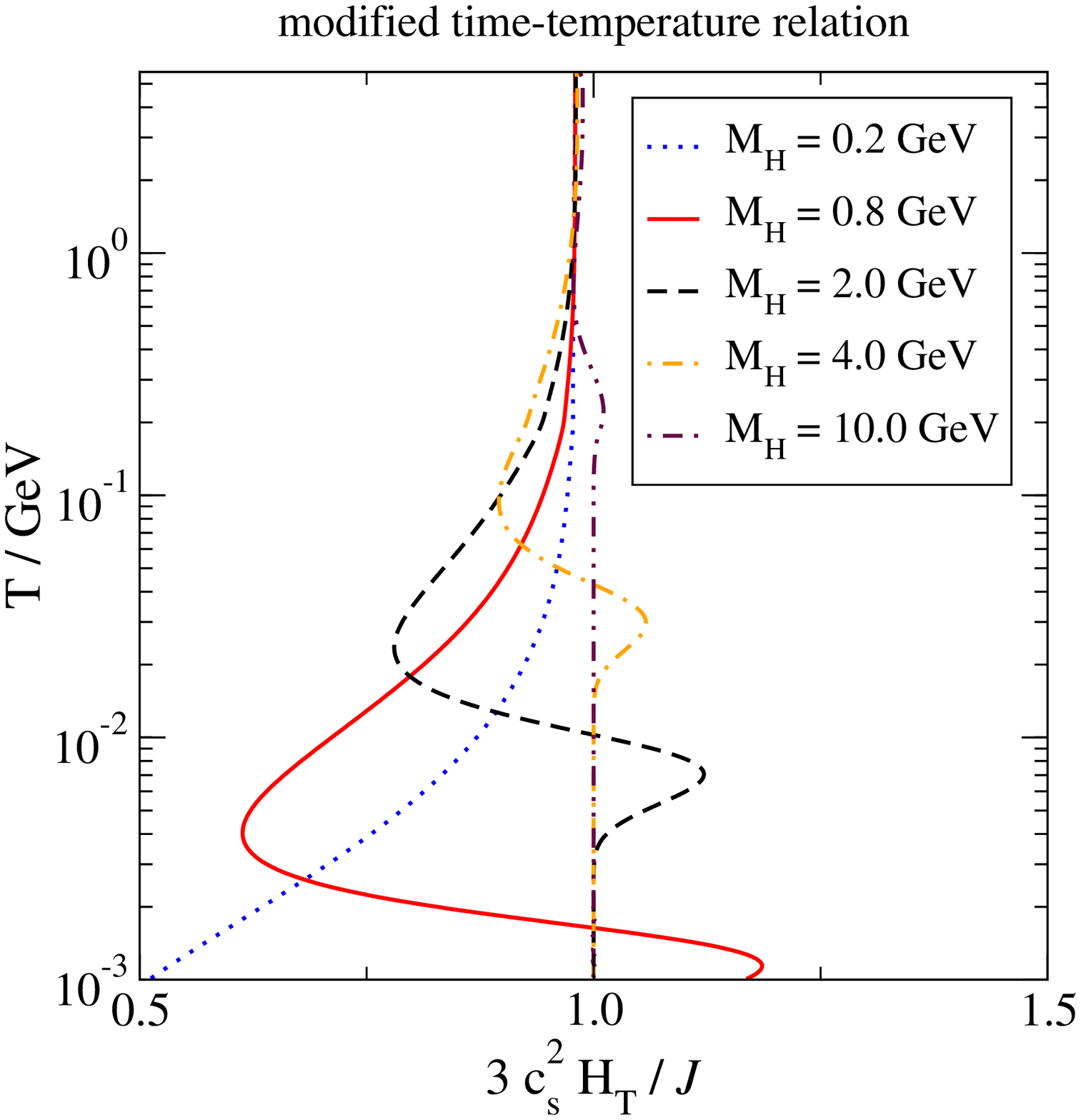}%
}

\caption[a]{\small
 Left: entropy release due to out-of-equilibrium decays of 
 heavy right-handed neutrinos. For $M^{ }_{\sH} = 0.2$~GeV the 
 process takes place at temperatures lower than those shown. 
 Right: the (inverse) Jacobian from \eq\nref{J}, normalized to 
 the value within the Standard Model~\cite{eos}, with
 $
 H^{ }_\T \equiv \sqrt{{8\pi e^{ }_\T }/{( 3 m^{2}_\rmi{Pl} )}} 
 $. Two competing effects in $\mathcal{J}$ lead to a 
 non-monotonous behaviour: an increase of $H$, 
 and a decrease due to the non-equilibrium term
 placed just after $3H$ in \eq\nref{J}. 
}

\la{fig:entropy}
\end{figure}

Numerical solutions for $s^{ }_\T a^3$, obtained from 
\eqs\nref{xeq_1} and \nref{xeq_2}, are shown in \fig\ref{fig:entropy}(left). 
In \fig\ref{fig:entropy}(right) we show the corresponding $\mathcal{J}$
from \eq\nref{J}, normalized to its Standard Model value. 
According to \fig\ref{fig:entropy}(left), 
entropy release substantially reduces any yields that were 
generated at $T > 0.1$~GeV, if $M^{ }_{\sH} \;\lsim\; 2$~GeV.\footnote{%
 This statement depends somewhat on the values of the Yukawas
 chosen, cf.\ \se\ref{se:params} and item (ii) in \se\ref{se:summary}.
 } 
In contrast 
the effect from $\mathcal{J}$ is moderate, if dark matter 
production is peaked at $T > 0.1$~GeV.

Even if only the evolution of $Y^+_{\I\I}$ is directly coupled with
the evolutions of $s^{ }_\T a^3$ and $\mathcal{J}$
(cf.\ \eqs\nref{xeq_1}, \nref{xeq_2}), the results in 
\fig\ref{fig:entropy} also influence other evolution equations.  
The redshift factor from \eq\nref{kt} can be expressed as 
\be
 \frac{a(t^{ }_0)}{a(t)}
 = 
 \biggl\{ \frac{s(T)}{s(T^{ }_0)} \biggr\}^{\fr13}
 \exp
 \biggl\{ \fr13 \Bigl[
    \ln(s^{ }_\T a^3)(t^{ }_0) - \ln(s^{ }_\T a^3)(t) 
                \Bigr] 
 \biggr\}
 \;. 
\ee
As we have 
replaced $t$ as an integration variable through $x = \ln(T^{ }_\rmi{max}/T)$
with the help of $\mathcal{J}$,
the co-moving momentum 
will from now on be denoted by $\kT^{ }$.
Defining $Y \equiv n/s^{ }_\T$, evolution equations for particle densities
and phase space distributions from \se\ref{se:review} are transcribed as  
\ba
 \dot{n}(t) = 
 \mathcal{F}
 & \;\Longrightarrow\; &
 \mathcal{D}^{ }_x Y = 
 \frac{ \mathcal{F} }{\mathcal{J} s^{ }_\T }
 \;, \quad
 \mathcal{D}^{ }_x \; \equiv \; 
 \partial^{ }_x + 
 \partial^{ }_x\ln(s^{ }_\T a^3) 
 \;, \la{d_exp_Y} \\ 
 \dot{f}(t,k) 
 = \mathcal{G}(k) 
 & \;\Longrightarrow\; & 
 \partial^{ }_x f(x,\kT^{ }) 
 = \frac{ \mathcal{G}(\kT^{ }) }{\mathcal{J}} 
 \;. \la{d_exp_f} 
\ea

%
\section{Simplified treatment of heavy flavours}
\la{se:heavy}

It was indicated in the previous section that for the heavy
flavours it is advantageous to resort to momentum averaging. 
Moreover, it is convenient to go over
into an interaction picture. 
In order to simplify the notation of 
\eqs\nref{d_exp_Y} and \nref{d_exp_f}, 
let us denote 
\be
 (...)' \;\equiv\; \mathcal{D}^{ }_x (...)
 \;, \quad
 \widehat{Q} \;\equiv\; \frac{Q}{\mathcal{J}}
 \;. 
\ee
Then the role of the free Hamiltonian is played by
$
    \diag\langle \widehat{\omega}^{ }_2,\widehat{\omega}^{ }_3 \rangle^{ }_1
    - \bigl\langle \widehat{H}^{+}_{\!\sH} \bigr\rangle^{ }_1
$,
where the averaging 
$
 \langle ... \rangle^{ }_1
$ 
is defined in \eq\nref{momav_1}. From here 
we can subtract the trace part without loss of generality. 
The remaining upper diagonal appearing in 
\eq\nref{d_rhoH_1p2} is defined as 
\be
 \widehat{H}^{ }_\rmi{fast} 
 \; \equiv \; 
 \frac{
  \langle \widehat{\omega}^{ }_2 - \widehat{\omega}^{ }_3 \rangle^{ }_1
  - \sum_a 
 [\, \phi^+_{(a)22} - \phi^+_{(a)33} \,]
 \, \langle \widehat{U}^+_{(a)\sH} \rangle^{ }_1 
 }
 {2}
 \;. \la{Hfast}
\ee
After the change of a picture, 
these diagonals do not appear on the right-hand side of the 
equations, whereas all non-diagonal coefficient functions get 
modified, as 
$
 (...)^{ }_{23} \to (...)^{ }_{23} \, (U^*)^2 
$, 
$
 (...)^{ }_{32} \to (...)^{ }_{32} \, (U)^2 
$, 
where the phase factor $U$ satisfies 
$
 U' = i \widehat{H}^{ }_\rmi{fast} U 
$.

In spite of the near-degeneracy of $M^{ }_2$ and $M^{ }_3$,
$\widehat{H}^{ }_\rmi{fast}$ defined in \eq\nref{Hfast}
becomes large at low temperatures (recall that $\widehat{\omega}^{ }_\I$
are normalized by $\mathcal{J}$, which decreases like
the Hubble rate, as $\propto T^2$).
In this situation the fast oscillations between the
heavy sterile neutrinos, induced by $\widehat{H}^{ }_\rmi{fast}$,  
can be ``integrated out''. 
Working to leading order in $1/\widehat{H}^{ }_\rmi{fast}$ as 
described in ref.~\cite{cpnumerics}, we find that in this regime
\ba
 Y'_a - \frac{Y'_\rmii{$B$}}{3}
 & \simeq & 
 \frac{4}{s^{ }_\T} \int_{\kT} \Bigl\{ 
   \bigl[f^{+}_{ } - \nF^{ }(\omega^{ }_1) \bigr]
   \, \widehat{B}^{+}_{(a)11} 
 + f^{-}_{ } \widehat{B}^{-}_{(a)11}
 - \nF^{ }(\omega^{ }_1) \bigl[1 - \nF^{ }(\omega^{ }_1)\bigr]\, 
  \widehat{A}^+_{(a)11}
 \Bigl\} 
 \nn[2mm]  
 & + &   
 {4} \; \biggl\{\;  
 {\textstyle\sum_{\I}}\, \phi^+_{(a)\I\I}
 \, 
 \Bigl[
  \bigl( Y^+_{\I\I} - Y^+_\rmi{eq} \bigr)\, 
  \bigl\langle \,\widehat{\!\bar{Q}}^{+}_{(a)\sH} \bigr\rangle^{ }_1 
 + Y^-_{\I\I}\, 
  \bigl\langle \widehat{Q}^{-}_{(a)\sH} \bigr\rangle^{ }_1
 - \bmu^{ }_a \, 
  \bigl\langle \widehat{Q}^{+}_{(a)\sH} \bigr\rangle^{ }_2
 \Bigr]
 \nn[2mm] 
 & + &  
 \frac{\phi^+_{(a)23} \, 
 \sum_b
 i \phi^-_{(b)23}
 \, 
  \langle \widehat{Q}^{-}_{(a)\sH} \rangle^{ }_1
 \, 
  \langle \widehat{Q}^{-}_{(b)\sH} \rangle^{ }_1 
 \bigl(\, 
  2 Y^+_\rmi{eq} 
  -  Y^+_{22} - Y^+_{33} 
 \,\bigr)
 }{
  \widehat{H}^{ }_\rmi{fast} 
 } \nn 
 & - &  
 \frac{i \phi^-_{(a)23} \, 
 \sum_b
 \phi^+_{(b)23}
 \, 
  \langle \widehat{Q}^{+}_{(a)\sH} \rangle^{ }_1
 \, 
  \langle \widehat{Q}^{+}_{(b)\sH} \rangle^{ }_1 
 \, 
 \bigl(\, 
  2 Y^+_\rmi{eq}
  -  Y^+_{22} - Y^+_{33} 
 \,\bigr)
 }{
  \widehat{H}^{ }_\rmi{fast}  
 }
 \; \biggr\} 
 \;. \la{dY_decoh}
\ea
Here we have complemented 
the momentum average in \eq\nref{momav_1} through
\be
 \langle ... \rangle^{ }_2 
 \; \equiv \; 
 \frac{ \int_{\kT}
  (...) \, \nF{}(\omega^{ }_{\sH}) 
 \, [ 1 - \nF{}(\omega^{ }_{\sH}) ] }{s^{ }_\T} 
 \;.
\ee
The helicity-symmetric diagonal components of the density matrix
evolve according to \eq\nref{xeq_1}, whereas the other components obey
\ba
 (Y^{\pm}_{23})' & \simeq & 
 - Y^{\pm}_{23} \, 
 {\textstyle\sum_{a,\I}} \,
  \phi^+_{(a)\I\I} \,
  \bigl\langle \widehat{Q}^{+}_{(a)\sH} \bigr\rangle^{ }_1 
 \;, \la{drho23_dec} \\[2mm] 
 (Y^{-}_{\I\I})' & \simeq & 
 2\, {\textstyle\sum_a} \, \phi^+_{(a)\I\I}
 \Bigl[
  \bigl( 
  Y^+_\rmi{eq}
  - 
  Y^+_{\I\I})\, 
  \bigl\langle \,\widehat{\!\bar{Q}}^{-}_{(a)\sH} \bigr\rangle^{ }_1 
  - 
  Y^-_{\I\I}\, 
  \bigl\langle \widehat{Q}^{+}_{(a)\sH} \bigr\rangle^{ }_1
  +
  \bmu^{ }_a\, 
  \bigl\langle \widehat{Q}^{-}_{(a)\sH} \bigr\rangle^{ }_2
 \Bigr]
 \;. \hspace*{5mm} \la{drhom_dec}  
\ea

%
\section{Resonant contribution in light flavour}
\la{se:light}

The question arises whether momentum averaging could also be 
adopted for $f^{\pm}_{ }$. This is, however, hindered by the possible
appearance of a ``resonance'' in the coefficients 
$ 
 {Q}^{\pm}_{(a)\sL}, 
 \,{\!\bar{Q}}^{\pm}_{(a)\sL}, 
$
which parametrize the evolution of $f^{\pm}_{ }$ through
\eqs\nref{B_11}--\nref{D_11}. The resonance originates through the 
helicity-conserving indirect contribution, 
which for $M^{ }_1 \ll \kT^{ }$ has the form
\be
 Q^{ }_{(a-)\sL} +  
 \bar{Q}^{ }_{(a-)\sL} \Bigr|^\rmi{indirect}_{ }
 \; \approx \; 
 \frac{v^2 M_1^2 \Gamma^{ }_{\!u}}
 {2[(M_1^2 + 2 \omega^{ }_1 b)^2 + (\omega^{ }_1 \Gamma^{ }_{\! u})^2]}
 \;. \la{resonance}
\ee
The function $b$ has a C-even and C-odd part; the latter, which 
is proportional to chemical potentials, is denoted by 
$
 b |^{ }_{\mu} \equiv c
$.
At low temperatures the C-even part is to a good approximation 
proportional to $\omega^{ }_1$~\cite{nr}. Therefore we may write
\be
 b = \tilde{b}\, \omega^{ }_1 + c
 \;. 
\ee
The function $\tilde{b}$ is positive, whereas
$c$ is odd in the interchange $\mu^{ }_i\leftrightarrow -\mu^{ }_i$. 
Therefore, after extracting the C-even $Q^{ }_{(a-)}$
and the C-odd $\bar{Q}^{ }_{(a-)}$ from \eq\nref{resonance}
by symmetrizing and antisymmetrizing in chemical potentials, 
respectively, both contain one appearance of 
\be
 Q^\rmi{res}_{ }(\omega^{ }_1) \; \equiv \; 
 \frac{v^2 M_1^2}{2\omega^{ }_1}
 \frac{\omega^{ }_1 \Gamma^{ }_{\!u}}
 { \mathcal{F}^2
 + (\omega^{ }_1 \Gamma^{ }_{\! u})^2  }
 \;, \quad
 \mathcal{F} \; \equiv \; 
 2 \tilde{b}\, \omega_1^2 - 2 |c| \omega^{ }_1 + M_1^2 
 \;. \la{resonance2}
\ee
For small $\omega^{ }_1\Gamma^{ }_{\!u}$ this can be approximated as 
\be
  Q^\rmi{res}_{ }(\omega^{ }_1) \; \approx \; 
 \frac{v^2 M_1^2}{2\omega^{ }_1}\, \pi\delta(\mathcal{F})
 \;. \la{resonance3}
\ee
This is qualitatively different from non-resonant contributions, 
which are proportional to $\Gamma^{ }_{\!u}$. 

We observe from \eqs\nref{resonance2} and \nref{resonance3} 
that resonances exists if $c^2 > 2 \tilde{b} M_1^2$, 
and they are located at 
\be
 \omega^\rmi{res}_{\pm} = 
 \theta(c^2 - 2 \tilde{b} M_1^2) \, 
 \frac{|c| \pm \sqrt{c^2 - 2 \tilde{b} M_1^2}}{2\tilde{b}}
 \;. 
\ee
Recalling that
$
 \tilde{b} \simeq 80 G_\rmii{F}^2 T^4 
$
and 
$
 c \simeq -\mu^{ }_a G^{ }_\rmii{F} T^2 
$~\cite{nr}, 
where $ G^{ }_\rmii{F} $ is the Fermi constant, 
resonances are important if 
$
 |\mu^{ }_a| \gsim 10 M^{ }_1
$. 
For $M^{ }_1 = 7$~keV and $T\sim 0.2$~GeV, this requires
$|\bmu^{ }_a| \gg 10^{-4}$.  

%
\section{Parameter values and initial conditions}
\la{se:params}

We start by considering the  
benchmark point $\filledsquare$ from ref.~\cite{degenerate}, 
tuned to produce the observed baryon asymmetry as well as maximally
large low-temperature lepton asymmetries, within a specific slice
of the parameter space. The most important parameters are 
$M^{ }_{\sH} \approx 0.7732$~GeV, $\Delta M = 10^{-11}$~GeV, 
$\im z = -0.15$, where the last one refers to 
the Casas-Ibarra parameter fixing the absolute value of
the neutrino Yukawas~\cite{ci}. 
The small $|\im z|$ implies $|h^{ }_{\I a}| \simeq 2\times 10^{-8}$.
The corresponding active-sterile mixings, 
$
 {|h^{ }_{\I a}| v}/({\sqrt{2}M^{ }_{\I}})
 \simeq 4.5\times 10^{-6}
$, 
are tiny and thus
challenging to constrain in (future) experiments.

For the light sector we fix the overall active-sterile mixing
angle to a maximal suggested value~\cite{observe2}, 
$\sin^2(2\theta) \equiv \sum_a 2 |h^{ }_{1 a}|^2 v^2 / M_1^2 
\simeq 2\times 10^{-10}$, 
i.e.\ $\sum_a |h^{ }_{1 a}|^2 \simeq 8 \times 10^{-26}$
for $M^{ }_1 \simeq 7$~keV. 
Furthermore we set $h^{ }_{1e} = h^{ }_{1\mu} = h^{ }_{1\tau}$,
which according to the web site associated with 
ref.~\cite{dmpheno} leads to maximal 
efficiency in dark matter production. 
Thus, $|h^{ }_{1a}| \simeq 1.6 \times 10^{-13}$ for all $a$.

The initial conditions for the evolution are set at a temperature 
$T \approx 5$~GeV, where the system is to a good approximation
in a stationary state~\cite{degenerate}. 
Taking also into account that rate coefficients are dominated by 
helicity-conserving contributions at low temperatures, 
\eqs\nref{dY_decoh}--\nref{drhom_dec} imply that
\be
 Y^{+}_{\I\I} \big|^{ }_{T\,\approx\, 5~{\rm GeV}}
 \;\approx\; 
 Y^+_\rmi{eq}
 \;, \quad
 Y^{-}_{\I\I} \big|^{ }_{T\,\approx\, 5~{\rm GeV}}
 \;\approx\;
 - \bmu^{ }_\rmi{ave} \, X^-_\rmi{eq} 
 \;, \la{initial}
\ee
where 
$
 \bmu^{ }_\rmi{ave} \equiv \frac{1}{3}\sum_a \bmu^{ }_a
$
and 
$
 X^-_\rmi{eq} \, \equiv \, 
 \langle 1 \rangle^{ }_2
$.
To be optimistic, we multiply 
lepton asymmetries obtained in ref.~\cite{degenerate}
by a factor two, leading to the initial condition 
$Y^{ }_a - \YB^{ }/3 \simeq - 6.2 \times 10^{-7}$ 
for all $a$, 
which fixes the chemical potential appearing in \eq\nref{initial}
as $\bmu^{ }_\rmi{ave} \simeq -6.6 \times 10^{-5}$. 
The initial baryon asymmetry is set at the observed value 
$\YB^{ } = 0.87 \times 10^{-10}$; in view of entropy dilution, it should
be taken to be somewhat larger at the beginning, 
but this has little effect on our considerations here, and is also
easy to achieve in practice~\cite{degenerate}.

In addition to the benchmark point $\filledsquare$, 
we have carried out further scans like in 
ref.~\cite{degenerate}, and again multiplied the 
corresponding lepton asymmetries by a factor two.
This leads to two further parameter points which serve to 
illustrate the dependence on $M^{ }_{\sH}$. 
Initial lepton asymmetries can be kept large 
by decreasing $M^{ }_{\sH}$, but there is not much room here, given that
according to refs.~\cite{Neff0,Neff} 
there is a cosmological lower bound $M^{ }_{\sH} \gsim 0.1$~GeV.
We have chosen $M^{ }_{\sH} = 0.2$~GeV as a lighter mass;  
as we will see, this is already problematic
(the other parameters are $\Delta M = 10^{-10}$~GeV, $\im z = -0.66$, 
$Y^{ }_a - \YB^{ }/3 = -5.0 \times 10^{-7}$). 
As a heavier mass we have settled on $M^{ }_{\sH} = 4.0$~GeV
($\Delta M = 10^{-14}$~GeV, $\im z = -0.20$, 
$Y^{ }_a - \YB^{ }/3 = -5.4 \times 10^{-11}$), which clearly 
illustrates how results depend on $M^{ }_{\sH}$. We have also 
carried out further runs with $M^{ }_{\sH} = 2.0, 10.0$~GeV and 
these confirm the overall picture. 

%
\section{Numerical solution}
\la{se:abundance}

Important ingredients characterizing the solution of the rate equations 
are equilibration rates, which determine how efficiently 
different components of the density
matrix approach their would-be equilibrium values. As an example, 
consider the dimensionless combination appearing in \eq\nref{drho23_dec}
but for simplicity normalized to the thermal Hubble rate rather
than~$\mathcal{J}$, 
\be
 \tilde{\Gamma}^{ }_{\sH} \; \equiv \; 
  \frac { \langle \Gamma^{ }_{\sH} \rangle^{ }_1 }{3 c_s^2 H^{ }_{\T}} 
 \;, \quad 
 \Gamma^{ }_{\sH} \; \equiv \; 
 \sum_{a,\I}
 \, \phi^+_{(a)\I\I} \, 
 {Q}^{+}_{(a)\sH}
 \;. \la{equilrate_H}
\ee
The result is shown in \fig\ref{fig:equilrate}(left). We observe that 
the system can follow equilibrium 
(i.e.\ that 
$
 \tilde{\Gamma}^{ }_{\sH} \; \gsim \; 1
$) 
when $T \; \gsim \; 2 $~GeV, 
but at $T < 2$~GeV there is a period when 
this should not happen. 
At very low temperatures, rates are dominated by vacuum decays,
and the system again approaches equilibrium. 

\begin{figure}[t]

\hspace*{-0.1cm}
\centerline{%
 \epsfysize=7.5cm\epsfbox{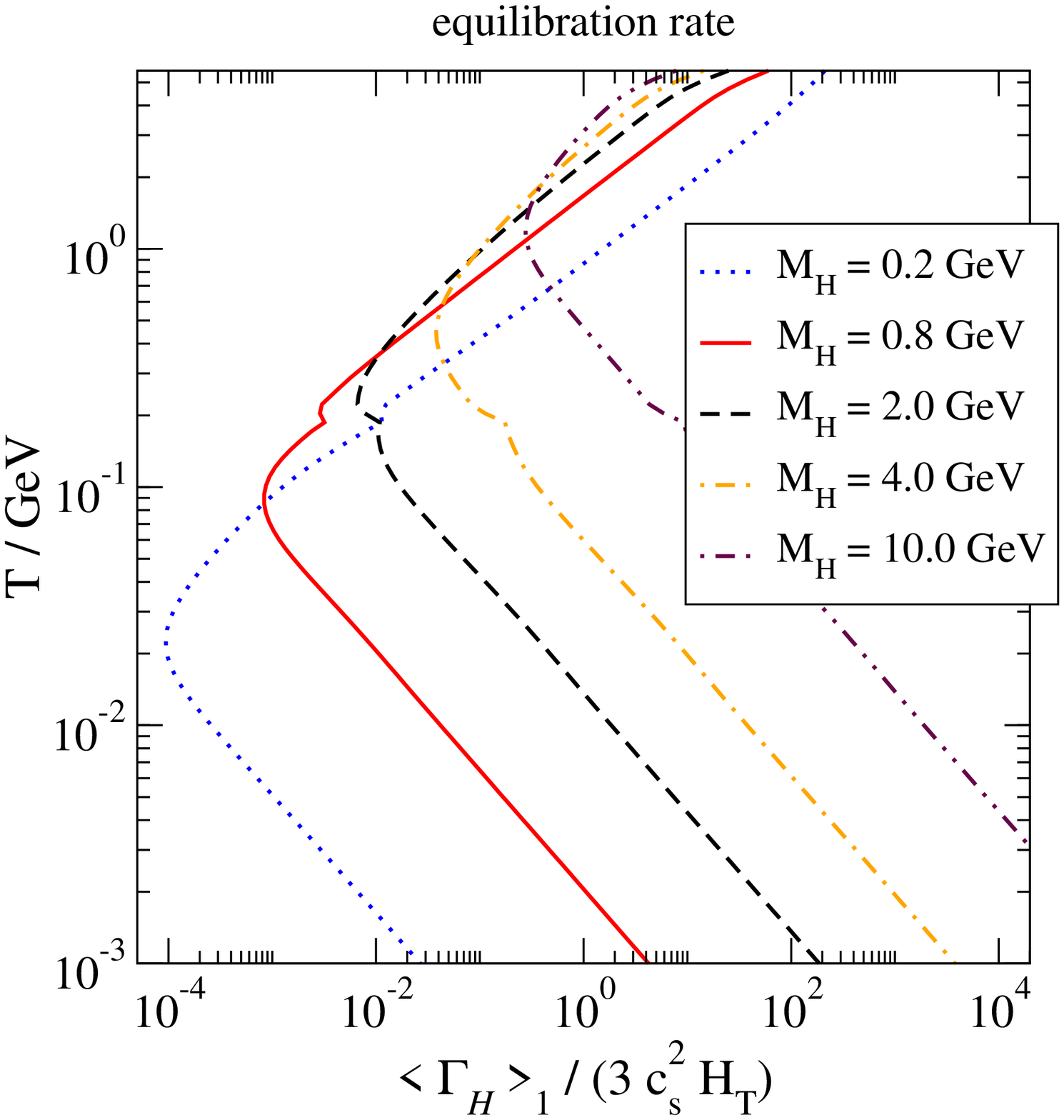}%
 \hspace{0.5cm}%
 \epsfysize=7.5cm\epsfbox{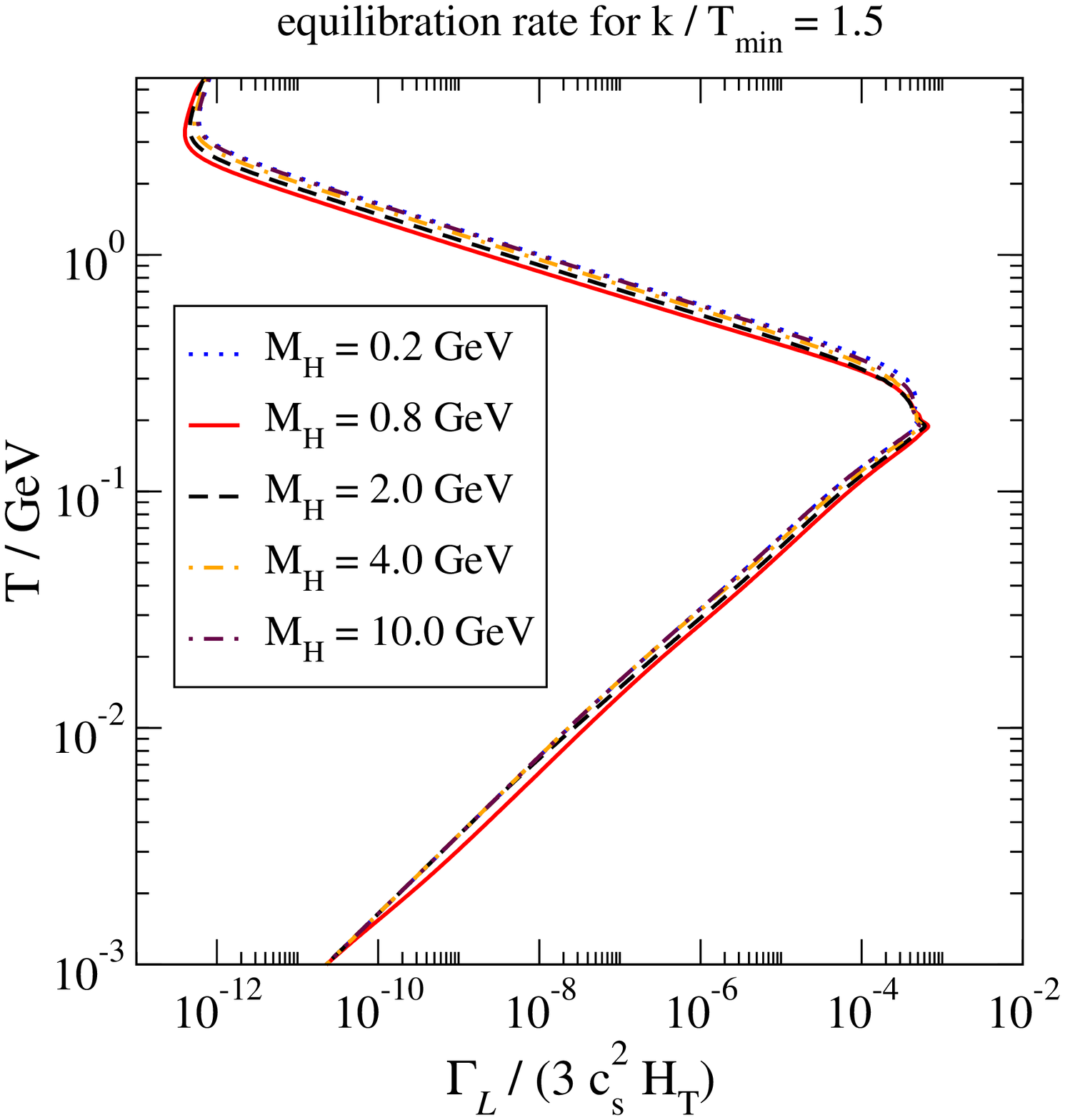}%
}

\caption[a]{\small
 Left: an equilibration rate of heavy flavours, defined 
 in \eq\nref{equilrate_H}.
 The small spike just below $T = 0.2$~GeV
 originates from a dip in $c_s^2$
 around the QCD crossover 
 (we employ the parametrization from ref.~\cite{eos}). 
 Right: an equilibration rate of light flavours, 
 for a specific comoving momentum mode.  
 For this plot, 
 the yields $Y^{ }_a$ and $\YB^{ }$ have been kept
 fixed at their initial values.
}

\la{fig:equilrate}
\end{figure}

For the light flavour, we 
show an equilibration rate 
from \eqs\nref{d_rhoL_1p2}, \nref{D_11}, 
evaluated at a fixed comoving momentum, 
in \fig\ref{fig:equilrate}(right)
($
 \Gamma^{ }_{\sL} \equiv
 2 \sum_a \phi^+_{(a)11} \, 
 {Q}^{+}_{(a)\sL} 
$). 
Given that in the full range
$
 \tilde{\Gamma}^{ }_{\sL} \equiv 
 \Gamma^{ }_{\sL} / ( {3 c_s^2 H^{ }_{\T}})
 \ll 10^{-3}
$, 
the light flavour never comes near thermal equilibrium. 

\begin{figure}[t]

\hspace*{-0.1cm}
\centerline{%
 \epsfysize=5.0cm\epsfbox{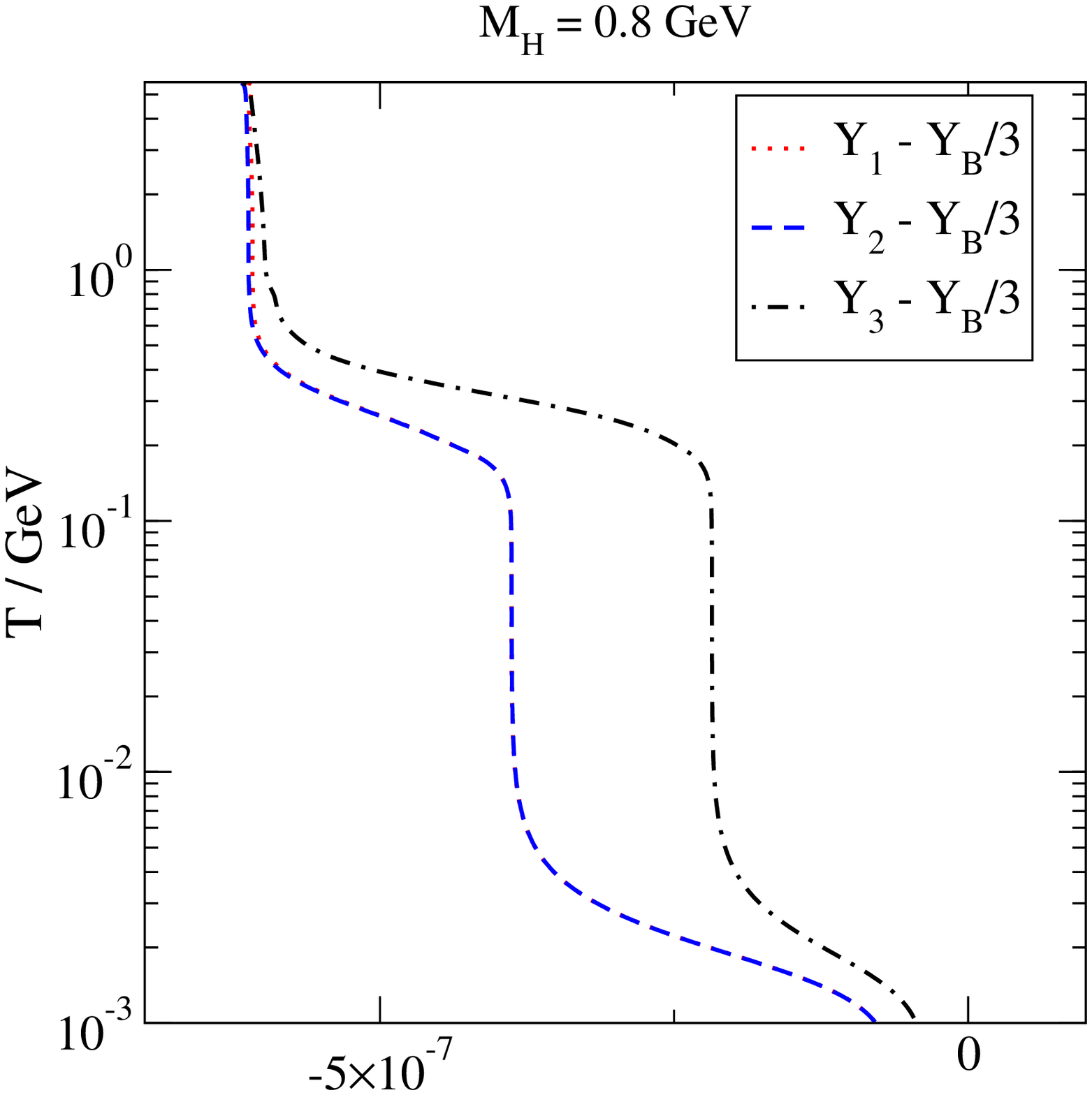}%
 \hspace{0.5cm}%
 \epsfysize=5.0cm\epsfbox{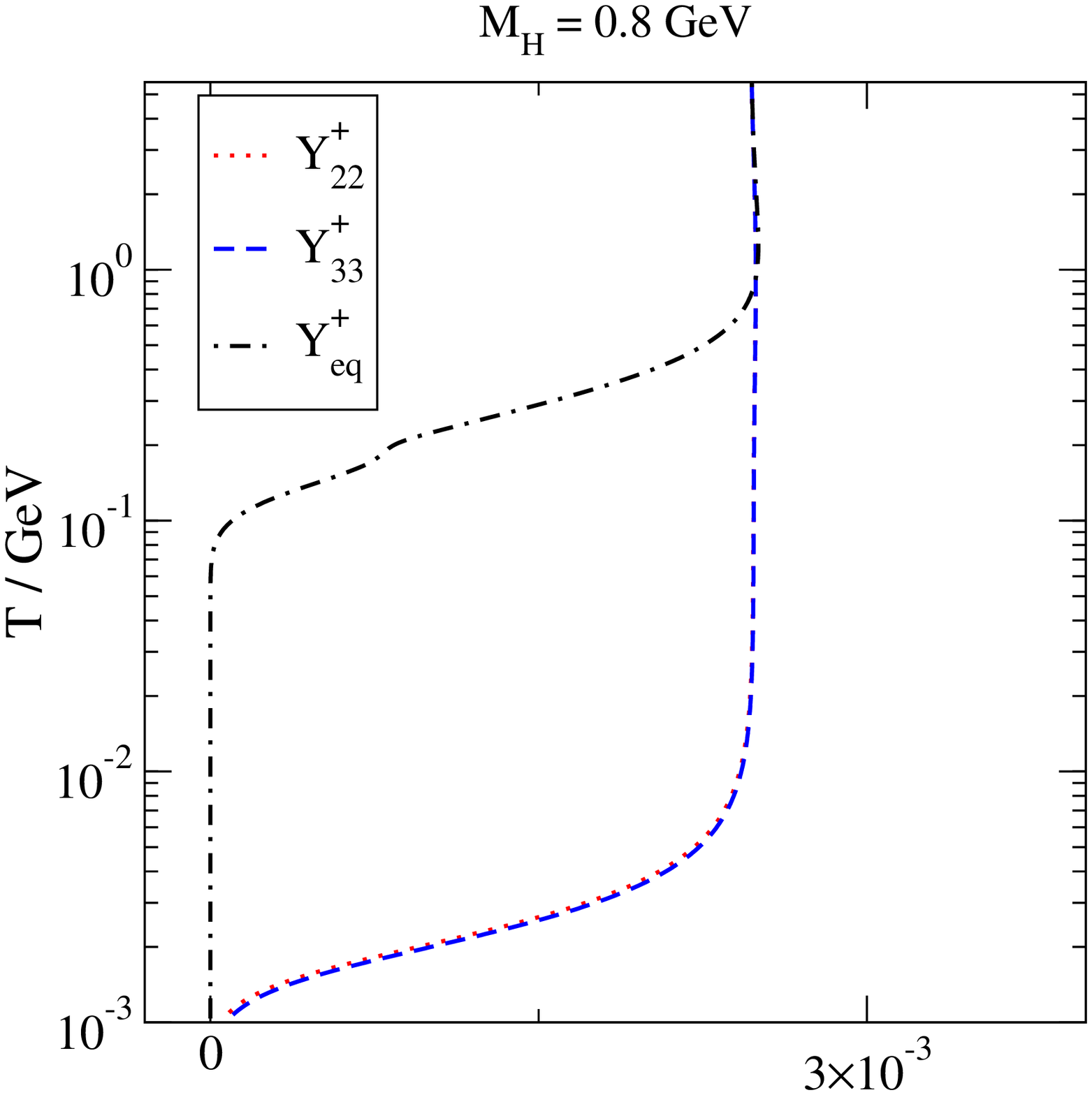}
 \hspace{0.5cm}%
 \epsfysize=5.0cm\epsfbox{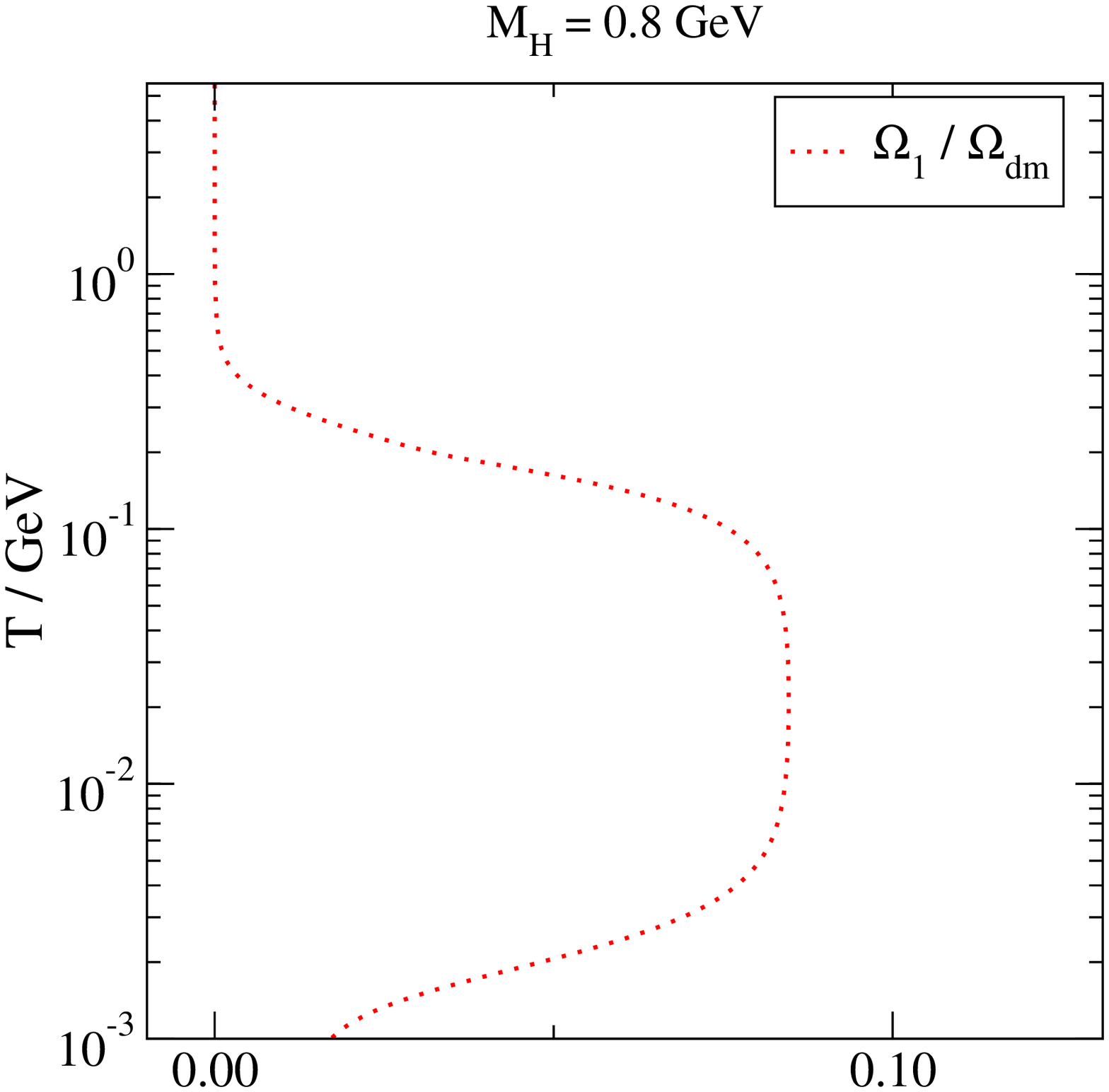}
}

\caption[a]{\small
 Left: 
 lepton asymmetries, $Y^{ }_a - {\YB^{ }}/{3}$, 
 at $M^{ }_{\sH} \approx 0.8$~GeV. 
 The decrease at $T = 0.1...0.3$~GeV is caused by 
 conversion into dark matter sterile neutrinos. 
 Middle: 
 helicity-symmetric components of the density matrix, 
 compared with the equilibrium value 
 $Y^+_\rmii{eq}$. 
 Right: 
 the fraction of dark matter that
 $Y^{+}_{11} \equiv \int_{\vec{k}^{ }_\T} f^+_{ } / s^{ }_{\T}$
 accounts for, 
 cf.\ \eq\nref{fracOmegadm}.
 The decrease at low $T$ is due to entropy dilution. 
}

\la{fig:Y_H}
\end{figure}

For benchmark $\filledsquare$ (i.e.\ $M^{ }_{\sH} \approx 0.8$~GeV), 
the solutions of the rate equations for lepton asymmetries and 
the density matrix of the heavy sector 
are shown in \fig\ref{fig:Y_H}. At first the density matrix follows
the equilibrium form, but at $T \ll 1$~GeV the equilibrium form
starts to decrease as mass effects 
become important. The actual 
solution cannot immediately 
follow this change, given that the equilibration
rate has become small. 

\begin{figure}[t]

\hspace*{-0.1cm}
\centerline{%
 \epsfysize=5.0cm\epsfbox{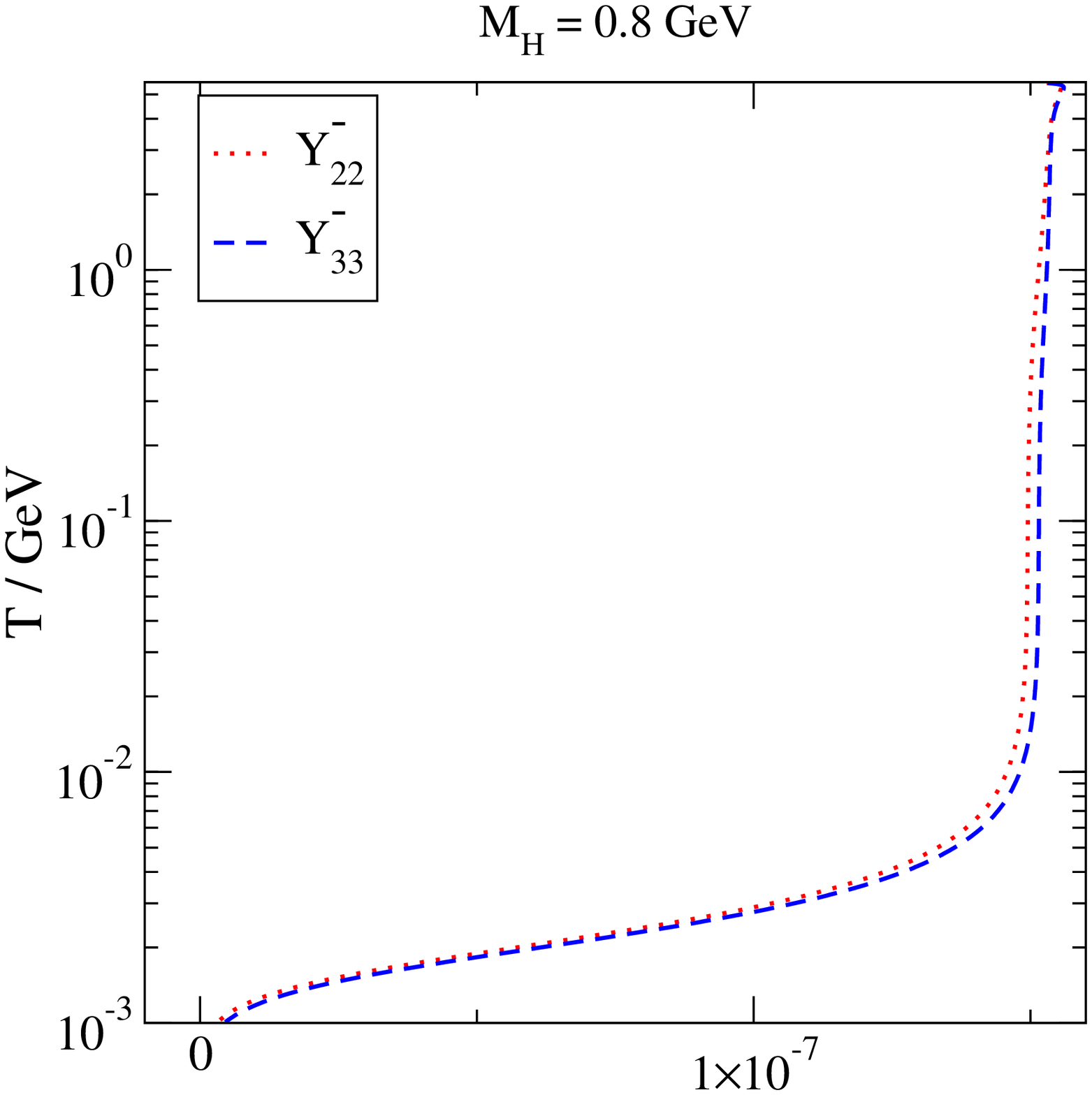}
 \hspace{0.5cm}%
 \epsfysize=5.0cm\epsfbox{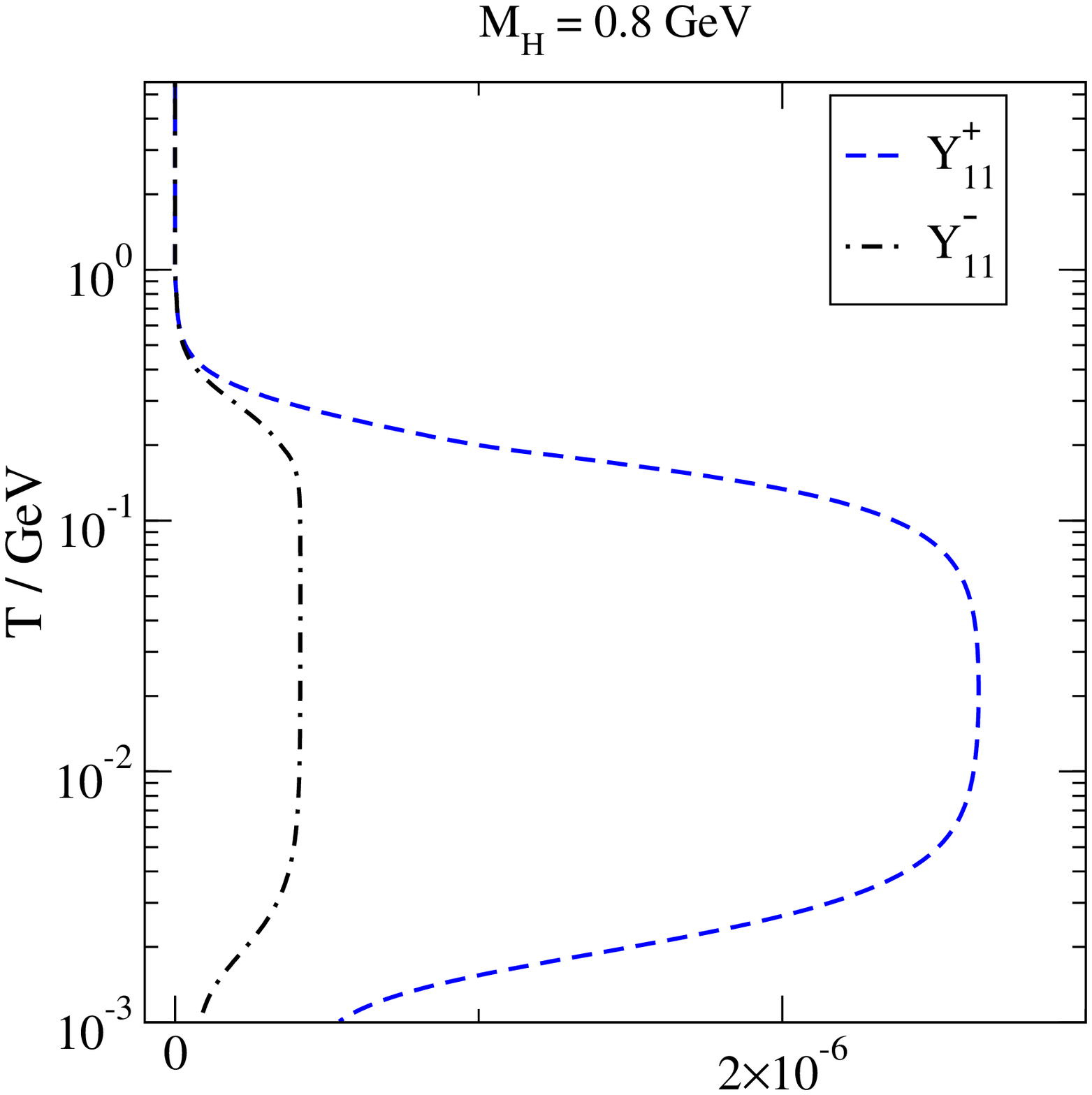}
 \hspace{0.5cm}%
 \epsfysize=5.0cm\epsfbox{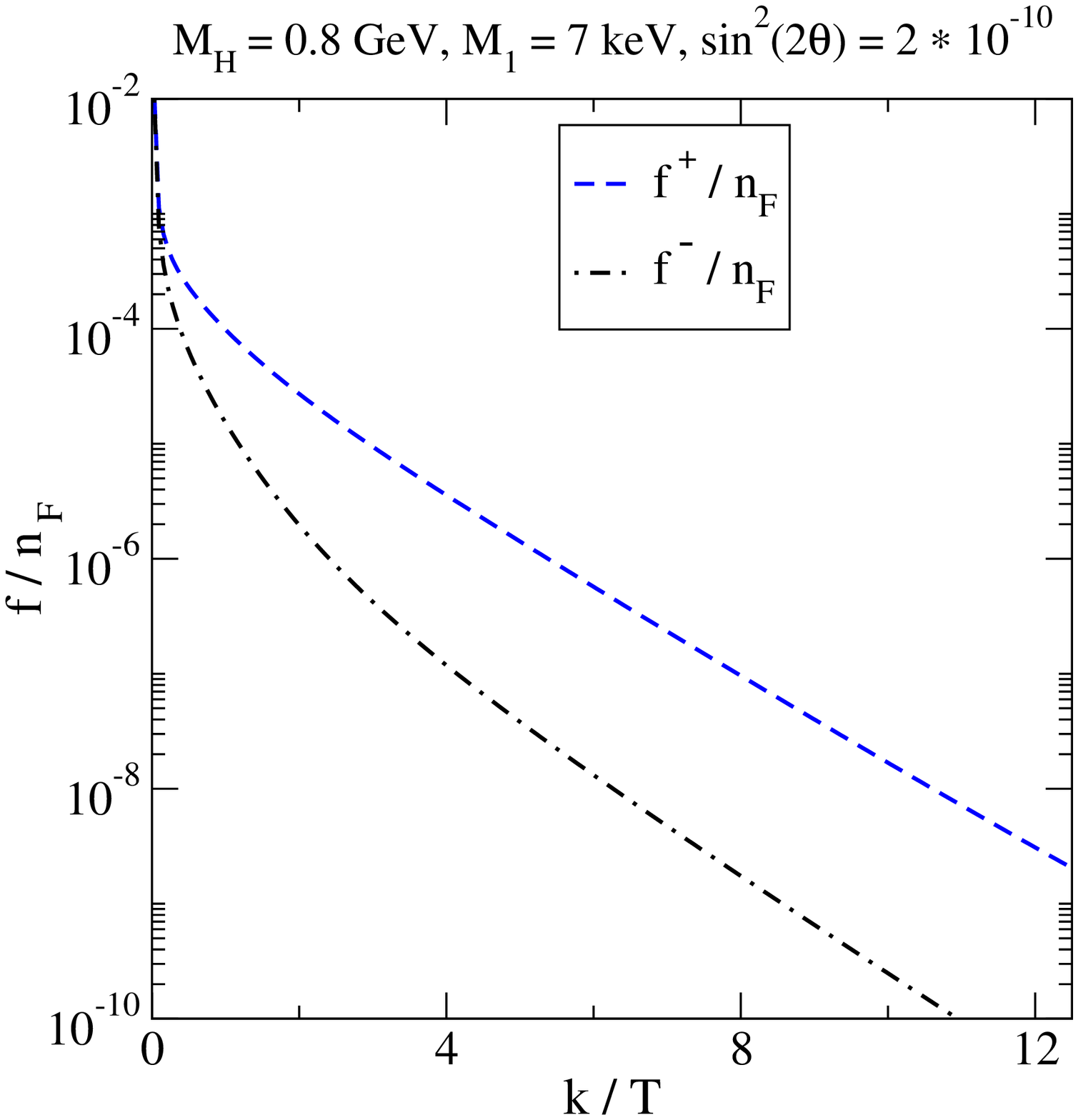}%
}

\caption[a]{\small
 Left: yields 
 associated with the helicity-asymmetric components of 
 the heavy right-handed neutrinos, 
 for $M^{ }_{\sH} \approx 0.8$~GeV. 
 Middle: helicity symmetries and asymmetries of the light flavour, 
 $Y^{\pm}_{11} \equiv \int_{\vec{k}^{ }_\T} f^{\pm}_{ } / s^{ }_{\T}$.
 Right:  
 The final dark matter spectra, $f^{\pm}_{ }$, 
 normalized to the Fermi distribution. 
}

\la{fig:various}
\end{figure}

We note from \fig\ref{fig:Y_H} that 
even if the density matrix deviates from equilibrium at low 
temperatures, there is no substantial re-generation of lepton asymmetries
taking place in this regime. The reason is that the  
rate coefficients are so small that the source terms, 
cf.\ the last lines of \eq\nref{dY_decoh}, remain inefficient. 

Making us of $\Omega^{ }_\rmi{dm} h^2 \approx 0.12$~\cite{planck} 
and $\rho^{ }_\rmi{cr} / [h^2 s(T^{ }_0)] = 3.65$~eV~\cite{pdg}, 
where $s(T^{ }_0)$
is the current entropy density, the fraction of dark matter carried by
the lightest right-handed neutrinos can be expressed as
\be
 \frac{ \Omega^{ }_{1} }{\Omega^{ }_\rmi{dm}}
 \approx 
 4.57\times Y^{+}_{11} \times \frac{M^{ }_1}{\mbox{eV}} 
 \;. \la{fracOmegadm}
\ee
We observe from \fig\ref{fig:Y_H}(right) that intermittently about
8.5\% of the total dark matter abundance could be accounted for, before
entropy dilution kicks in at late times. 

The yields of the helicity asymmetries are illustrated in 
\fig\ref{fig:various}. Helicity asymmetries remain modest
($Y^-_{\I\I} \ll Y^+_{\I\I}$, $I \in \{1,2,3\}$), 
which is a manifestation of the fact that thermal production dominates
over resonant production (one helicity state is produced from neutrinos,
the other from antineutrinos). 
The dark matter phase space spectra, which are
strongly tilted towards the IR compared with kinetically
equilibrated fermions, are shown in \fig\ref{fig:various}(right). 

\begin{figure}[t]

\hspace*{-0.1cm}
\centerline{%
 \epsfysize=5.0cm\epsfbox{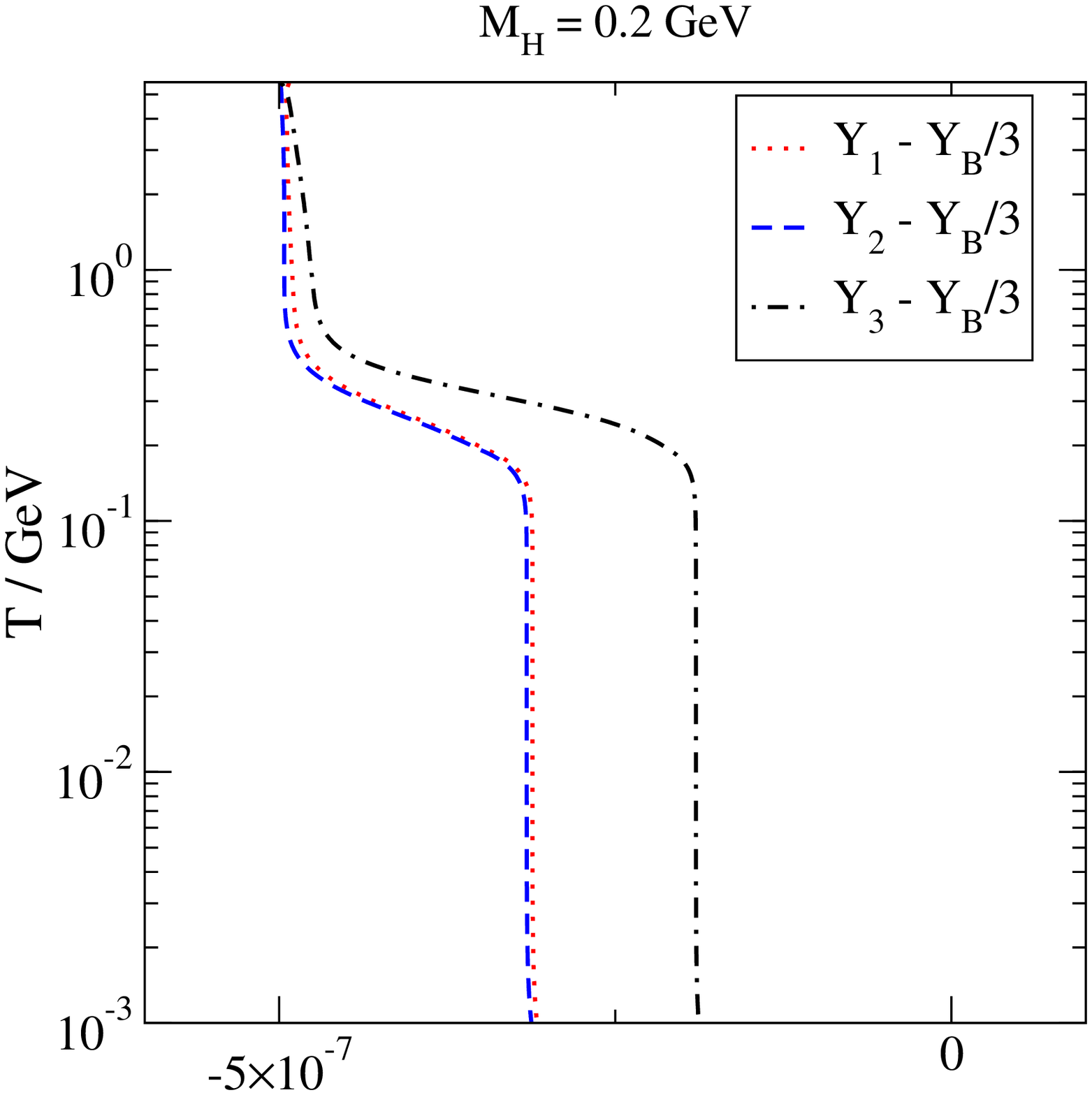}%
 \hspace{0.5cm}%
 \epsfysize=5.0cm\epsfbox{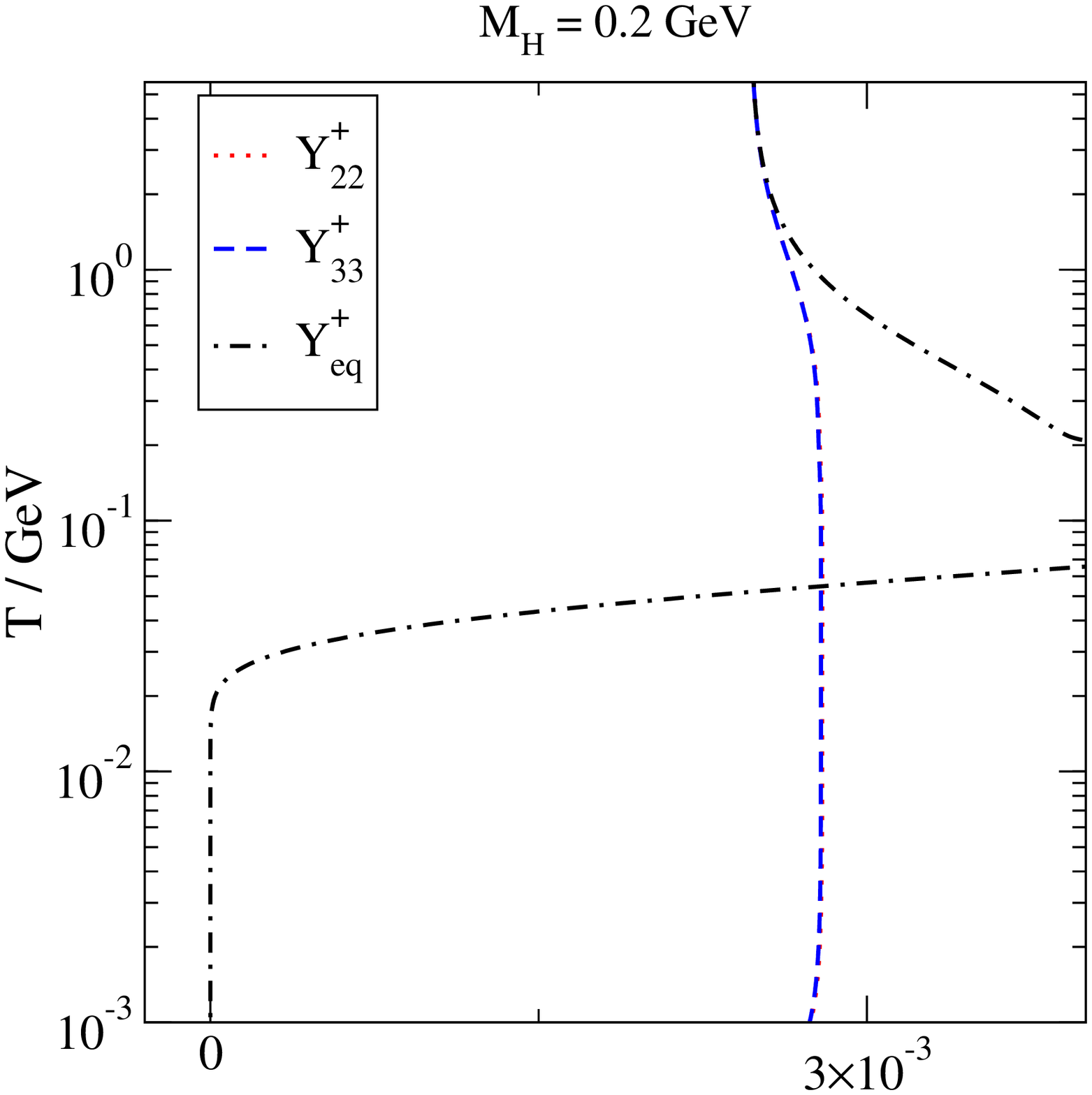}
 \hspace{0.5cm}%
 \epsfysize=5.0cm\epsfbox{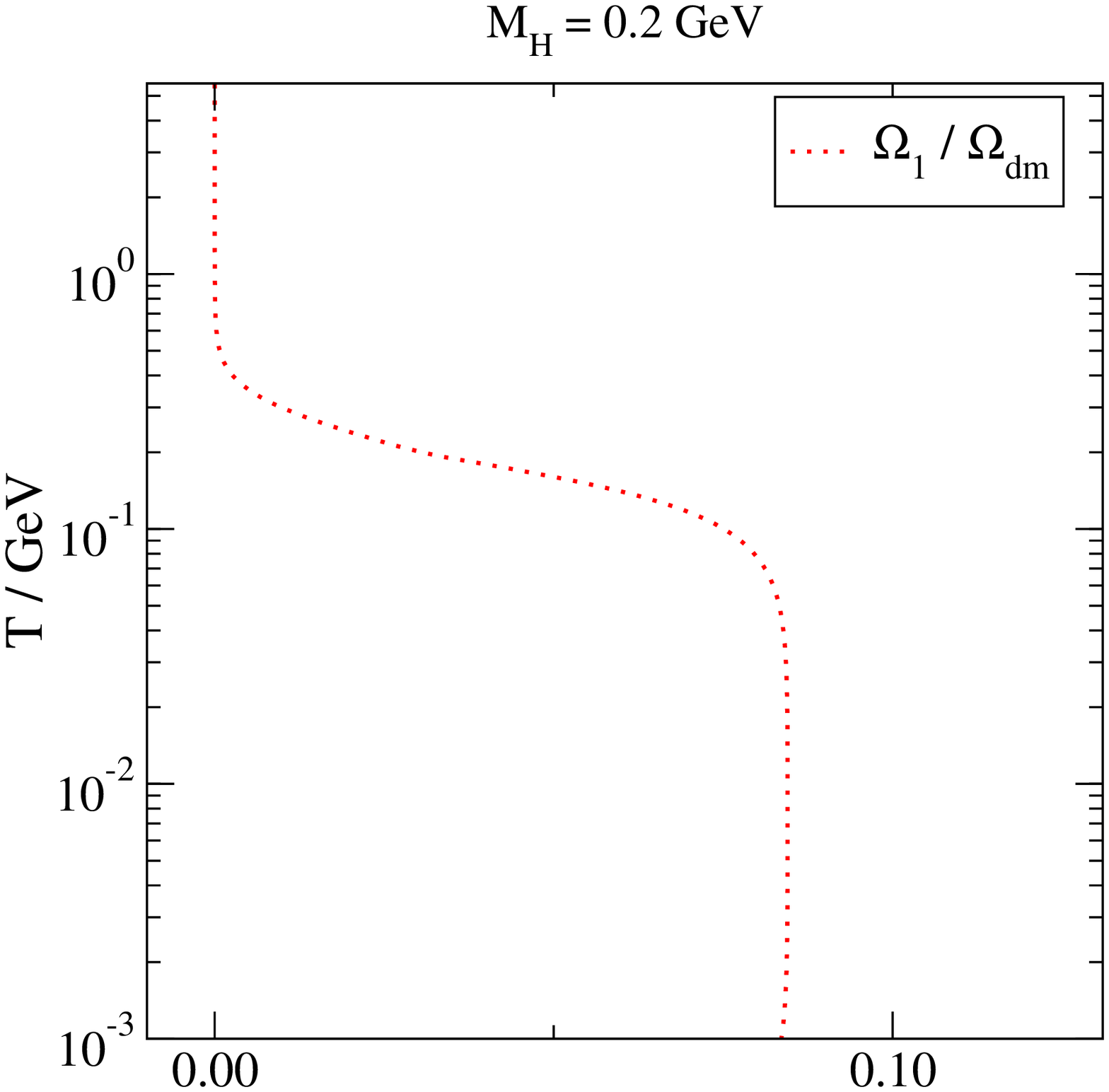}
}

\caption[a]{\small
 Like \fig\ref{fig:Y_H} but for $M^{ }_{\sH} = 0.2$~GeV
 (other parameters are listed around the end of \se\ref{se:params}). 
 Due to a smaller mass, the right-handed neutrinos do not
 decay efficiently, which leads
 to the problem that they may carry too much energy density 
 at late times (cf.,\ e.g.,\ refs.~\cite{Neff0,Neff}).
 Dark matter abundance is in the same ballpark as in 
 \fig\ref{fig:Y_H}, however entropy dilution has not started yet.
}

\la{fig:Y_Hm}
\end{figure}

\begin{figure}[t]

\hspace*{-0.1cm}
\centerline{%
 \epsfysize=5.0cm\epsfbox{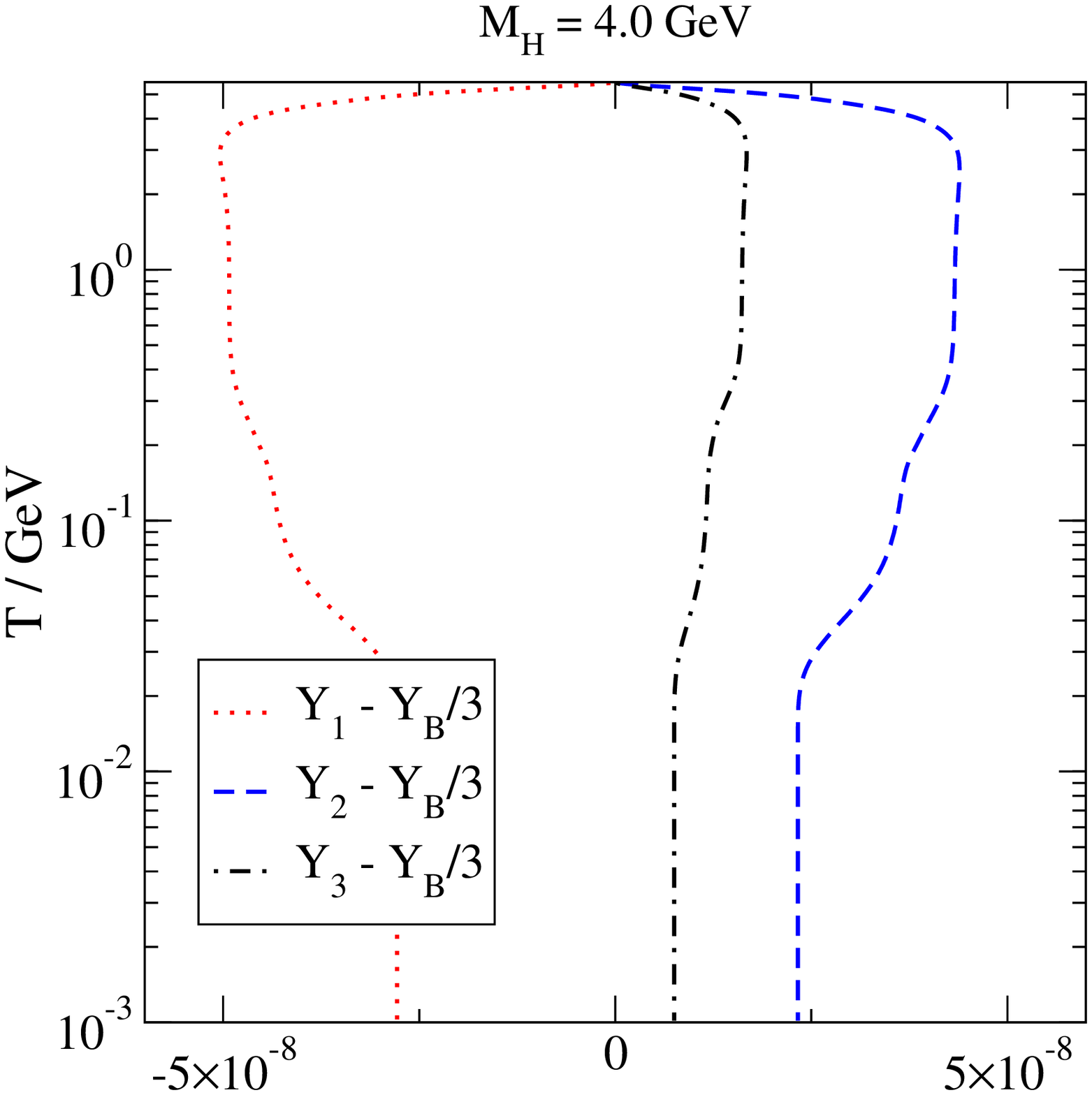}%
 \hspace{0.5cm}%
 \epsfysize=5.0cm\epsfbox{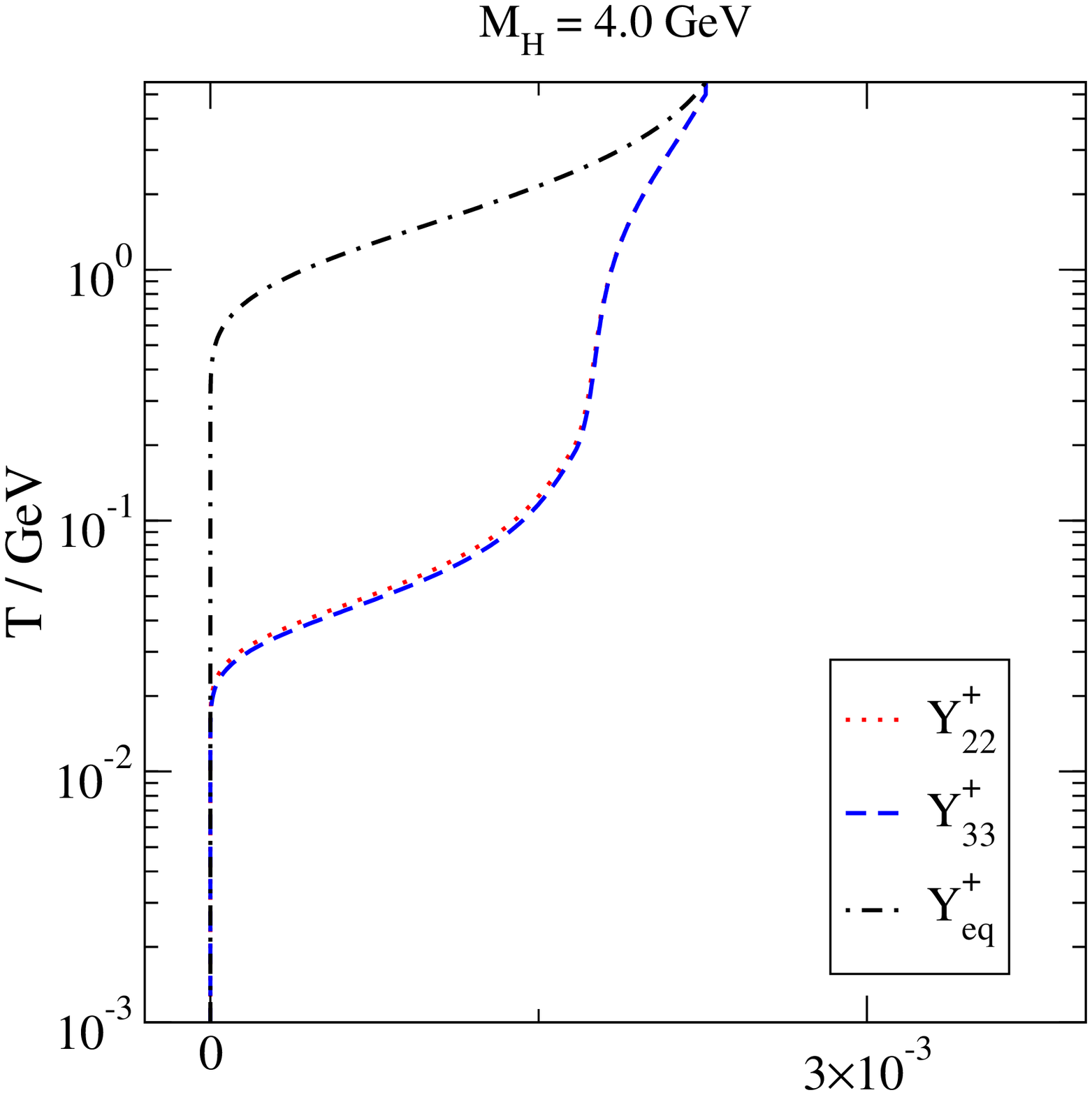}
 \hspace{0.5cm}%
 \epsfysize=5.0cm\epsfbox{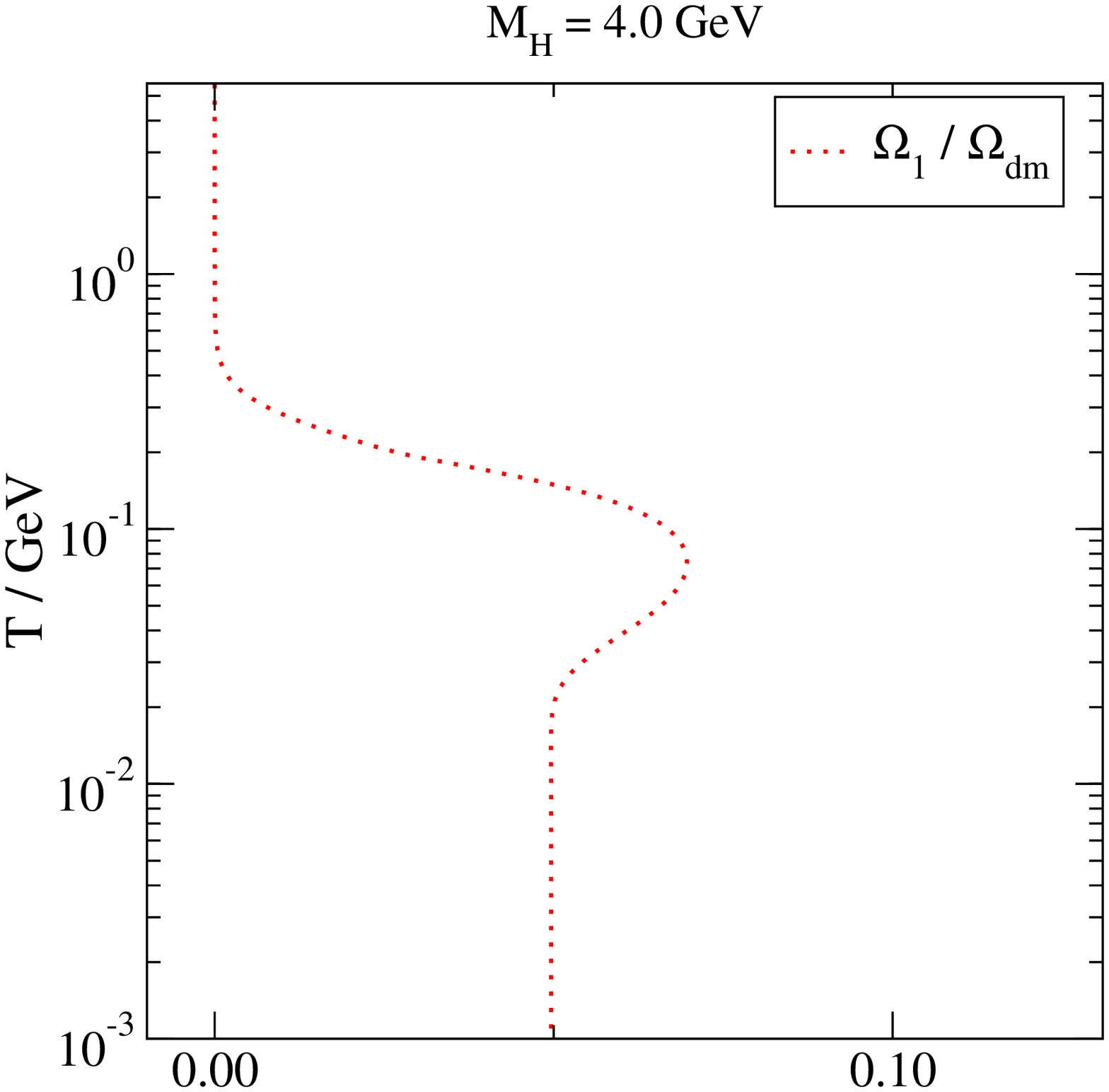}
}

\caption[a]{\small
 Like \fig\ref{fig:Y_H} but for $M^{ }_{\sH} = 4.0$~GeV
 (other parameters are listed around the end of \se\ref{se:params}). 
 In this case the initial lepton asymmetries obtained
 {\it \`a la} ref.~\cite{degenerate} are small,
 but novel asymmetries are generated  
 while $Y^+_{22}, Y^+_{33}$ are out of equilibrium
 (the suppression by $\widehat{H}^{ }_\rmi{fast}$ 
 in \eq\nref{dY_decoh} is originally moderate in this case, $\sim 1/500$). 
 However there is not much effect on the dark
 matter abundance. 
}

\la{fig:Y_Hp}
\end{figure}

Finally we consider the dependence of the final dark matter abundance on
the parameters of the heavy sector. 
Results for $M^{ }_{\sH} = 0.2$~GeV
are shown in \fig\ref{fig:Y_Hm}, 
and for $M^{ }_{\sH} = 4.0$~GeV
in \fig\ref{fig:Y_Hp}.
Despite a large variation in the original lepton asymmetries
and a re-generation of new ones in the latter case,  
the only important effect for dark matter abundance are 
the variations in the expansion history of the universe,  
as shown in \fig\ref{fig:entropy}.

%
\section{Summary and outlook}
\la{se:summary}

\begin{figure}[t]

\hspace*{-0.1cm}
\centerline{%
 \epsfysize=7.5cm\epsfbox{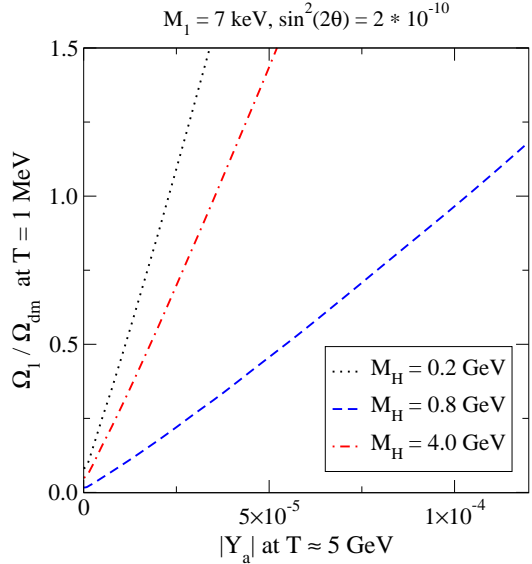}
}

\caption[a]{\small
 The ratio $\Omega^{ }_1 / \Omega^{ }_\rmi{dm}$
 at $T = 1$~MeV as a function of 
 the initial lepton asymmetry. Some further entropy
 dilution is expected at $T < 1$~MeV, particularly
 for $M^{ }_{\sH} = 0.2$~GeV, but the cosmological background is
 simultaneously becoming more complicated,
 as active neutrinos decouple and 
 big bang nucleosynthesis starts. In any case, 
 obtaining  $\Omega^{ }_1 \approx \Omega^{ }_\rmi{dm}$ would require
 lepton asymmetries about two orders of magnitude larger than 
 those found in ref.~\cite{degenerate}. 
}

\la{fig:fracOmegadm_scan}
\end{figure}

The purpose of this paper has been to update our previous sterile 
neutrino dark matter analysis~\cite{dmpheno} by fixing the initial 
lepton asymmetries to maximal values that can be produced by the dynamics 
of GeV-scale right-handed neutrinos~\cite{degenerate}. The parameters of 
the latter are constrained to be responsible 
for generating the active neutrino masses and mixing angles~\cite{ci}. 
We permit for the generation of further lepton asymmetries 
in the low-temperature decays of the  
right-handed neutrinos,\footnote{%
  However, the parametric fine tunings described in \se{2.6.2} of 
  ref.~\cite{late}, requiring a specific choice of CP phases,
  have not been imposed.} 
by including
both the light and heavy sterile flavours in the set of rate equations, 
and track the modification of the universe expansion caused by
the energy density carried and entropy released by the heavy flavours.
In addition we resolve both helicity states of the sterile neutrinos;
this is important for heavy flavours given that initial lepton 
asymmetries are correlated with helicity asymmetries~\cite{eijima,degenerate},
and for the light flavour given that resonant production~\cite{sf}
affects one helicity state only. 

Even though we do find rich dynamics in the heavy sector
(cf.\ \figs\ref{fig:Y_H}, \ref{fig:Y_Hm}, \ref{fig:Y_Hp}), the dark 
matter abundance does not vary greatly between the cases, reaching
typically less than 10\% of the observed value. The reason 
for this behaviour can be understood as follows. 
The dependence of $Y^+_{11}$ 
on lepton asymmetries must be quadratic
at small $Y^{ }_a$, given that energy density
is a C-even quantity.  
A strongly growing dependence only sets in at 
$|Y^{ }_a| \gg 10^{-6}$, cf.\ \fig\ref{fig:fracOmegadm_scan}.
This is associated with the dominance of resonant production, 
which in the language of \eq\nref{resonance3} requires  
$
 c^2 > 2 \tilde{b} M_1^2
$.
Given that the asymmetries obtained in ref.~\cite{degenerate} are
below this level,
dark matter production takes place 
predominantly through normal thermal processes. Therefore, our dark matter 
results are rather insensitive to the heavy sector, 
apart from its influence through the expansion of the universe, 
as depicted in \fig\ref{fig:entropy}. In order to account for 100\%
of dark matter, 
initial lepton asymmetries should be a factor $\sim 10^2$
larger than those found in ref.~\cite{degenerate}, 
i.e.\ of the order $|Y^{ }_a| \sim (2...10)\times 10^{-5}$, 
depending on entropy dilution.  

Even if we have failed to account for all of dark matter
through the dynamics of keV...GeV scale sterile neutrinos, 
the results could change in the future, for several reasons
(in line with the basic minimalistic premise of our study,  
we keep the Lagrangian of \eq\nref{L} intact for this discussion,
without any extra non-SM fields, 
and assume a standard cosmology): 
\bi

\item[(i)]
As the dark matter production is largely thermal rather than resonant, 
it is proportional to the rate coefficient $\Gamma^{ }_{\!u}$, which contains 
large hadronic uncertainties~\cite{hadronic}. It would be 
interesting to estimate or constrain $\Gamma^{ }_{\!u}$ 
through lattice simulations. 

\item[(ii)]
We have chosen the Yukawa couplings 
of the heavy flavours to be as small as possible, 
in order to diminish lepton number washout and therefore to have 
maximal initial asymmetries~\cite{degenerate}. 
However, as the initial asymmetries have little influence in any case, 
the Yukawas could be made larger, without spoiling
baryogenesis~
(cf.,\ e.g.,\ refs.~\cite{ph,inar,degenerate}). 
Then the heavy flavours
would stay closer to equilibrium and decay faster, producing less
entropy. Even though we do not expect substantial variations
of the dark matter abundance from here, a comprehensive study 
of the heavy flavour Yukawas would be welcome. This should 
also include
the search for potential ``atypical'' CP phases where late-time
lepton asymmetries might be anomalously large.  

\item[(iii)]
We have restricted ourselves to the 
SHiP window, $M^{ }_{\sH}< M^{ }_{B}\simeq 5$~GeV~\cite{ship}, 
but nature may have chosen otherwise. 
Increasing $M^{ }_{\sH}$ in the analysis of
ref.~\cite{degenerate}, we find that 
initial lepton asymmetries would be smaller then. 
However, as anticipated in refs.~\cite{singlet,late}, 
novel lepton asymmetries are produced later on
(cf.\ \fig\ref{fig:Y_Hp}). 
Therefore it seems promising to explore what happens with
larger values of $M^{ }_{\sH}$. Then, however,  
$2\leftrightarrow 2$ scatterings entering the rate coefficients 
need to be addressed 
without resorting to the approximation $M^{ }_{\sH} \ll \mW^{ }$, 
which poses a significant technical challenge. The
initial temperature should be chosen in the regime $T \gg M^{ }_{\sH}$,
i.e.\ larger than here. 

\item[(iv)]
Additional semi-conserved quantities such as chiral charges 
or helical magnetic fields have long been speculated to play 
a role in cosmology~(cf.,\ e.g.,\ 
refs.~\cite{eR1,eR2,Bext1,Bext2,Bext3,Bext4}), 
and they could conceivably interfere with 
late-time lepton asymmetries as well.  

\item[(v)]
Last but not least, the observational status 
of the dark matter sterile neutrino remains unclear. Here
we have relied on the indications in refs.~\cite{observe1,observe2}, 
however other parameter values could be studied 
within our framework, and might change the conclusions. 

\ei

To summarize, we have established a framework which permits to study
sterile neutrino dark matter production in a non-degenerate 1+2 flavour
situation, with ongoing lepton number violation and with the 
heavy flavours falling out of equilibrium and gradually decaying. 
As a proof of concept, we have excluded several SHiP-like 
benchmarks as an explanation for all of dark matter. Broader 
parameter scans may help to bridge the gap. 

%
\section*{Acknowledgements}

We thank M.~Shaposhnikov for encouraging us to verify, several years ago, 
that sterile neutrino dark matter production is most efficient when all 
$|h^{ }_{1a}|$ are equally large; 
these tests are documented on
the web page associated with ref.~\cite{dmpheno}.
We are grateful to S.~Eijima,  
M.~Shaposhnikov and I.~Timiryasov for helpful suggestions.
M.L.\ was partly supported by the Swiss National Science Foundation
(SNF) under grant 200020-168988.

%
\appendix
\renewcommand{\thesection}{Appendix~\Alph{section}}
\renewcommand{\thesubsection}{\Alph{section}.\arabic{subsection}}
\renewcommand{\theequation}{\Alph{section}.\arabic{equation}}

%


\begin{thebibliography}{99}

\bibitem{dw}
  S.~Dodelson and L.M.~Widrow,
  {\it Sterile-neutrinos as dark matter,}
  Phys.\ Rev.\ Lett.\  {72} (1994) 17
  [hep-ph/9303287].

\bibitem{sf}
  X.-D.~Shi and G.M.~Fuller,
  {\it A New dark matter candidate: Nonthermal sterile neutrinos,}
  Phys.\ Rev.\ Lett.\  {82} (1999) 2832
  [astro-ph/9810076].

\bibitem{review}
  M.~Drewes {\it et al.},
  {\it A White Paper on keV Sterile Neutrino Dark Matter,}
  JCAP {01} (2017) 025
  [1602.04816].

\bibitem{singlet}
  M.~Shaposhnikov,
  {\it The $\nu$MSM, leptonic asymmetries, and properties of singlet fermions,}
  JHEP {08} (2008) 008
  [0804.4542].

\bibitem{ars}
  E.K.~Akhmedov, V.A.~Rubakov and A.Y.~Smirnov,
  {\it Baryogenesis via neutrino oscillations,}
  Phys.\ Rev.\ Lett.\  {81} (1998) 1359
  [hep-ph/9803255].

\bibitem{as}
  T.~Asaka and M.~Shaposhnikov,
  {\it The $\nu$MSM, dark matter and baryon asymmetry of the universe,}
  Phys.\ Lett.\ B {620} (2005) 17
  [hep-ph/0505013].

\bibitem{shifuller}
  M.~Laine and M.~Shaposhnikov,
  {\it Sterile neutrino dark matter as a consequence 
  of $\nu$MSM-induced lepton asymmetry,}
  JCAP {06} (2008) 031
  [0804.4543].

\bibitem{aba}
  T.~Venumadhav, F.Y.~Cyr-Racine, K.N.~Abazajian and C.M.~Hirata,
  {\it Sterile neutrino dark matter:
  Weak interactions in the strong coupling epoch,}
  Phys.\ Rev.\ D {94} (2016) 043515
  [1507.06655].

\bibitem{dmpheno}
  J.~Ghiglieri and M.~Laine,
  {\it Improved determination of sterile neutrino dark matter spectrum,}
  JHEP {11} (2015) 171
  [1506.06752], 
  and the associated web site 
  {\tt http://www.laine.itp.unibe.ch/dmpheno/}. 

\bibitem{late}
  L.~Canetti, M.~Drewes, T.~Frossard and M.~Shaposhnikov,
  {\it Dark Matter, Baryogenesis and Neutrino Oscillations from
  Right Handed Neutrinos,}
  Phys.\ Rev.\ D {87} (2013) 093006
  [1208.4607].

\bibitem{degenerate}
  J.~Ghiglieri and M.~Laine,
  {\it Precision study of GeV-scale resonant leptogenesis,}
  JHEP {02} (2019) 014
  [1811.01971].

\bibitem{eijima}
  S.~Eijima and M.~Shaposhnikov,
  {\it Fermion number violating effects in low scale leptogenesis,}
  Phys.\ Lett.\ B {771} (2017) 288
  [1703.06085].

\bibitem{observe1}
  E.~Bulbul {\it et al},
  {\em Detection of An Unidentified Emission Line in the 
  Stacked X-ray spectrum of Galaxy Clusters,}
  Astrophys.\ J.\  {789} (2014) 13
  [1402.2301].

\bibitem{observe2}
  A.~Boyarsky {\it et al},
  {\it Unidentified Line in X-Ray Spectra of the Andromeda Galaxy
  and Perseus Galaxy Cluster,}
  Phys.\ Rev.\ Lett.\  {113} (2014) 251301
  [1402.4119].

\bibitem{sr}
  G.~Sigl and G.~Raffelt,
  {\it General kinetic description of relativistic mixed neutrinos,}
  Nucl.\ Phys.\ B {406} (1993) 423.

\bibitem{selfE}
  D.~B\"odeker, M.~Sangel and M.~W\"ormann,
  {\it Equilibration, particle production, and self-energy,}
  Phys.\ Rev.\ D {93} (2016) 045028
  [1510.06742].

\bibitem{cptheory}
  J.~Ghiglieri and M.~Laine,
  {\it GeV-scale hot sterile neutrino oscillations:
  a derivation of evolution equations,}
  JHEP {05} (2017) 132
  [1703.06087].

\bibitem{numsm} 
  T.~Asaka, M.~Laine and M.~Shaposhnikov,
  {\it Lightest sterile neutrino abundance within the $\nu$MSM,}
  JHEP {01} (2007) 091; 
  {\it ibid.} {02} (2015) 028 (E)
  [hep-ph/0612182], 
  and the associated web site 
  {\tt http://www.laine.itp.unibe.ch/neutrino-rate/}.

\bibitem{broken} 
  J.~Ghiglieri and M.~Laine,
  {\it Neutrino dynamics below the electroweak crossover,}
  JCAP {07} (2016) 015
  [1605.07720].

\bibitem{st}
  R.J.~Scherrer and M.S.~Turner,
  {\it Decaying particles do not ``heat up'' the Universe,}
  Phys.\ Rev.\ D {31} (1985) 681.
  
\bibitem{entropy}
  T.~Asaka, M.~Shaposhnikov and A.~Kusenko,
  {\it Opening a new window for warm dark matter,}
  Phys.\ Lett.\ B {638} (2006) 401
  [hep-ph/0602150].

\bibitem{kinetic} 
  T.~Asaka, S.~Eijima and H.~Ishida,
  {\it Kinetic Equations for Baryogenesis via Sterile Neutrino Oscillation,}
  JCAP {02} (2012) 021
  [1112.5565].

\bibitem{cpnumerics}
  J.~Ghiglieri and M.~Laine,
  {\it GeV-scale hot sterile neutrino oscillations: a numerical solution,}
  JHEP {02} (2018) 078
  [1711.08469].

\bibitem{eos}
  M.~Laine and Y.~Schr\"oder,
  {\it Quark mass thresholds in QCD thermodynamics,}
  Phys.\ Rev.\ D {73} (2006) 085009
  [hep-ph/0603048],
  and the associated web site 
  {\tt http://www.laine.itp.unibe.ch/eos06/}. 


\bibitem{nr}
  D.~N\"otzold and G.~Raffelt,
  {\it Neutrino Dispersion at Finite Temperature and Density,}
  Nucl.\ Phys.\ B {307} (1988) 924.

\bibitem{ci}
  J.A.~Casas and A.~Ibarra,
  {\it Oscillating neutrinos and $\mu \to e \gamma$,}
  Nucl.\ Phys.\ B {618} (2001) 171
  [hep-ph/0103065].

\bibitem{Neff0}
  O.~Ruchayskiy and A.~Ivashko,
  {\it Restrictions on the lifetime of sterile neutrinos from
  primordial nucleosynthesis,}
  JCAP {10} (2012) 014
  [1202.2841].

\bibitem{Neff}
  P.~Hern\'andez, M.~Kekic and J.~Lopez-Pavon,
  {\it $N_{\rm eff}$ in low-scale seesaw models versus
  the lightest neutrino mass,}
  Phys.\ Rev.\ D {90} (2014) 065033
  [1406.2961].

\bibitem{planck}
  N.~Aghanim {\it et al.} [Planck Collaboration],
  {\it Planck 2018 results. VI. Cosmological parameters,}
  1807.06209.

\bibitem{pdg}
  M.~Tanabashi {\it et al.} [Particle Data Group],
  {\it Review of Particle Physics,}
  Phys.\ Rev.\ D {98} (2018) 030001.

\bibitem{hadronic}
  T.~Asaka, M.~Laine and M.~Shaposhnikov,
  {\it On the hadronic contribution to sterile neutrino production,}
  JHEP {06} (2006) 053
  [hep-ph/0605209].

\bibitem{ph}
  P.~Hern\'andez, M.~Kekic, J.~L\'opez-Pav\'on, J.~Racker and J.~Salvado,
  {\it Testable Baryogenesis in Seesaw Models,}
  JHEP {08} (2016) 157
  [1606.06719].

\bibitem{inar}
  S.~Eijima, M.~Shaposhnikov and I.~Timiryasov,
  {\it Parameter space of baryogenesis in the $\nu$MSM,}
  JHEP {07} (2019) 077
  [1808.10833].

\bibitem{ship}
  S.~Alekhin {\it et al.},
  {\it A facility to Search for Hidden Particles at the CERN SPS:
  the SHiP physics case,}
  Rept.\ Prog.\ Phys.\  {79} (2016) 124201
  [1504.04855].

\bibitem{eR1}
  B.A.~Campbell, S.~Davidson, J.R.~Ellis and K.A.~Olive,
  {\it On the baryon, lepton flavor and right-handed electron asymmetries
  of the universe,}
  Phys.\ Lett.\ B {297} (1992) 118
  [hep-ph/9302221].

\bibitem{eR2}
  J.M.~Cline, K.~Kainulainen and K.A.~Olive,
  {\it Protecting the primordial baryon asymmetry from erasure by sphalerons,}
  Phys.\ Rev.\ D {49} (1994) 6394
  [hep-ph/9401208].

\bibitem{Bext1}
  M.~Joyce and M.E.~Shaposhnikov,
  {\it Primordial magnetic fields, right-handed electrons,
  and the Abelian anomaly,}
  Phys.\ Rev.\ Lett.\  {79} (1997) 1193
  [astro-ph/9703005].

\bibitem{Bext2}
  A.~Boyarsky, J.~Fr\"ohlich and O.~Ruchayskiy,
  {\it Self-consistent evolution of magnetic fields and
  chiral asymmetry in the early Universe,}
  Phys.\ Rev.\ Lett.\  {108} (2012) 031301
  [1109.3350].

\bibitem{Bext3}
  D.~B\"odeker and D.~Schr\"oder,
  {\it Equilibration of right-handed electrons,}
  JCAP {05} (2019) 010
  [1902.07220].

\bibitem{Bext4}
  D.G.~Figueroa, A.~Florio and M.~Shaposhnikov,
  {\it Chiral charge dynamics in Abelian gauge theories 
  at finite temperature,}
  1904.11892.

\end{thebibliography}
\end{document}